\documentclass[12pt]{iopart}

\usepackage{graphicx}
\usepackage{siunitx}
\usepackage[english]{babel}
\usepackage{amsfonts}
\usepackage{color}

\begin{document}

\title{Relaxation, chaos, and thermalization  in a  three-mode model of a BEC }

\author{M.A. Garcia-March$^1$, S. van Frank$^2$, M. Bonneau$^2$, J. Schmiedmayer$^2$, M. Lewenstein$^{1,3}$, and Lea F. Santos$^4$}
\address{$^1$ICFO -- Institut de Ciencies Fotoniques, The Barcelona Institute of Science and Technology, Av. Carl Friedrich Gauss 3, 08860 Castelldefels (Barcelona), Spain}
\address{$^2$Vienna Center for Quantum Science and Technology, Atominstitut, TU Wien, Stadionallee 2, 1020 Vienna, Austria}
\address{$^3$ICREA, Pg. Lluis Companys 23, 08010 Barcelona, Spain}
\address{$^4$Department of Physics, Yeshiva University,  New York, New York 10016, USA}

\date{\today}

\begin{abstract}
We study the complex quantum dynamics of a system of many interacting atoms in an elongated anharmonic trap. The system is initially in a Bose-Einstein condensed state, well described by Thomas-Fermi profile in the elongated direction and the ground state in the transverse  directions. After a sudden quench to a coherent superposition of the ground and  lowest energy  transverse modes, quantum dynamics starts.   We describe this process  employing a three-mode many-body model.  The experimental realization of this system displays decaying oscillations of the atomic density distribution.  While a mean-field description predicts perpetual oscillations of the atomic density distribution, our quantum many-body model exhibits a decay of the oscillations for sufficiently strong atomic interactions. We associate this decay with the fragmentation of the condensate during the evolution. The decay and fragmentation are also linked with the approach of the many-body model to the chaotic regime. The approach to chaos lifts degeneracies and increases the complexity of the eigenstates, enabling the relaxation to equilibrium and the onset of thermalization. We verify that the damping time and quantum signatures of chaos show similar dependences on the interaction strength and on the number of atoms.
\end{abstract}


\section{Introduction}
\label{sec:intro}

The emergence of new quantum simulators have allowed for a better understanding, description, and control of quantum many-body systems out of equilibrium. Progressively, approaches are found to explain and to take advantage of a variety of different factors that affect the dynamics of these complex systems, including range and strength of the interactions~\cite{Richerme2014,Jurcevic2014}, choice of initial state~\cite{Trotzky2008}, presence of disorder~\cite{Schreiber2015}, onset of quantum chaos~\cite{Torres2017PTR,Torres2018}, and proximity to critical points~\cite{Zurek2005}. Different behaviors have been identified at different time scales~\cite{Tavora2016,Tavora2017}, protocols to reach quantum speed limits have been engineered~\cite{Brouzos2015,vanFrank2016}, and conditions for isolated quantum systems to relax to equilibrium and to thermalize have been established~\cite{kinoshita06,Gogolin2016,Borgonovi2016,Alessio2016,Kaufman2016}. 

Relaxation in an isolated quantum system implies the approach of a set observables to a stationary value, deviations of which are very rare and negligible for long-time averages. The conditions required for relaxation to occur have been discussed in~\cite{Srednicki1996,Srednicki1999,Reimann2008,Linden2009,Linden2010,Short2011,Short2012,Reimann2012,Venuti2013,Zangara2013,Kiendl2017}. Thermalization is associated with the fact that the state of the system reached at long times cannot seemingly be distinguished from an {\it ad-hoc} defined thermal distribution~\cite{Gogolin2016,Borgonovi2016,Alessio2016,Eisert2015} ({\it ad-hoc} as it is defined for the particular closed system).

\begin{figure}[htb]
\centering
\includegraphics*[width=3.5in]{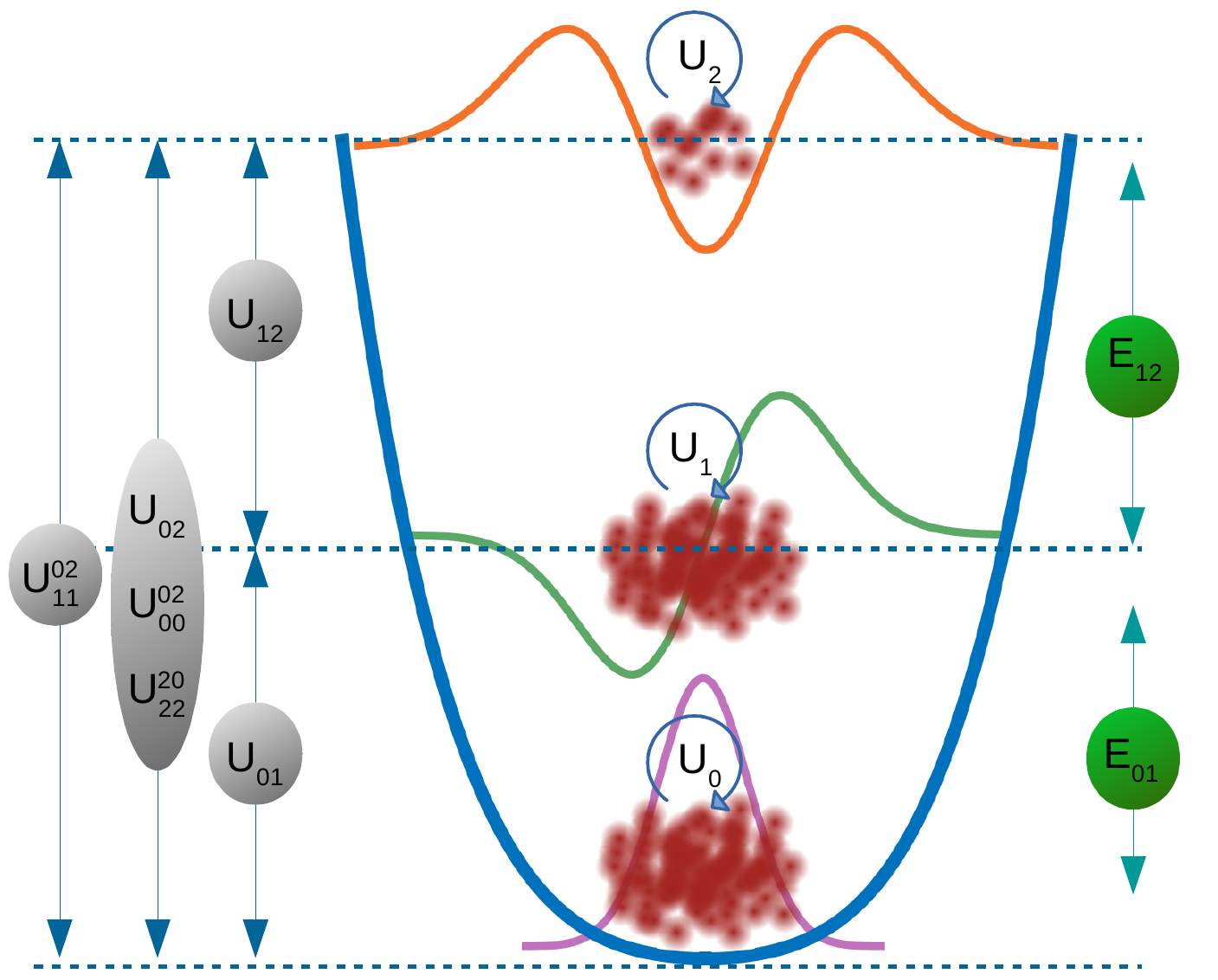}
\caption{ Illustration of the trapping potential and the Hamiltonian interaction parameters. $U_i$ represents the interaction coefficient between the atoms at level $i=0,1,2$ and $E_{ij}=E_j-E_i$ is the energy difference between the subsequent levels $i$ and $j$. The arrows between levels represent interaction  or transfer processes: (i) those with an interaction energy $U_{ij}$ are interactions between atoms at level $i$ and level $j$  or transfer of two atoms from level $i$ to $j$ or {\it vice versa}; (ii) those with an interaction energy $U_{ij}^{kl}$ destroy/create two atoms at level $i$ and $j$ and create/destroy one atom at level $i$ and two atoms at level $k$ and $l$.  Note that all transfers between the levels are led by the interactions.   }
\label{fig:3modes_pot}
\end{figure}

Experiments with cold atoms have a prominent place in studies of relaxation and thermalization due to their high level of  isolation~\cite{Greiner2002b,kinoshita06,hofferberth07,Strohmaier2007,Strohmaier2010,Cheneau2012,Gring2012,Trotzky2012,Langen2013,Kaufman2016}. Their access to precise coherence manipulation and to the preparation of desired initial states are essential for the investigation of quantum many-body dynamics~\cite{LangenARCMP}. In this context, a good example is a set of experiments with a quasi-one dimensional (quasi-1D) Bose-Einstein condensate (BEC) on an atom chip~\cite{Bucker2013,vanFrank2014,vanFrank2016}, which  successfully performed coherent transfer between motional states of the transverse  trapping  potential. This allowed for the  preparation of the condensate in a coherent superposition of the two lowest motional states, which was let to evolve in the trapping potential. 

The details of the dynamics of the above mentioned quasi-1D BEC are not yet entirely understood. At short times, the evolution of the initial superposition presents oscillations of the atomic density distribution~\cite{vanFrank2014}, which agrees with simulations based on a quasi-1D Gross-Pitaevskii equation (GPE). At longer time, the density distribution relaxes to a steady state~\cite{Bonneau2018}.  We note here that there are theoretical studies  on three coherent modes, which display many interesting phenomena~\cite{2004YukalovPRA,2004YukalovLP,2005YukalovLproc}. Nevertheless,  the damping of the oscillations is not captured by the mean-field (GPE) approximation with the physical parameters of the system. That is the GPE approach predicts perpetual oscillations of the atomic density distribution.
In this article, we investigate how a simple quantum many-body model can provide an explanation for the relaxation of this isolated system.  The model shows relaxation and thermalization, and hence it is a testbed for the analysis of theoretical bounds on relaxation times. 

Before discussing the quantum many-body model, we consider first a semiclassical model based on three modes. Similarly to the GPE, it accounts for the initial oscillations, but cannot explain their decay. A semiclassical two-mode model is even worse, being incapable of qualitatively describing the initial oscillations. The system investigated in~\cite{vanFrank2014,Bonneau2018} consisted of a degenerate gas of several hundreds $^{87}$Rb atoms, which justified the use of the GPE. Mean-field approximations effectively describe various phenomena in BEC and in many cases it is preferred over many-body approaches, which are computationally more involved and often intractable. However, mean-field approximations are by construction blind to the microscopic properties of individual atoms and do not account for collisions or quantum fluctuations. 

Our three-mode quantum many-body model initially prepared in the two lowest modes accounts for both the oscillations and their damping. As we show, the decay of the oscillations occurs as a pure quantum phenomenon and provides a neat example of relaxation in an isolated quantum system. By extrapolating the number of atoms reachable by our numerical tests to the number of atoms used in the experiment, we extract a value for the damping time. It is larger than the damping time in the experiment, but reproduces the decay of oscillations qualitatively.

Note that when referring to the decay of the oscillations, we employ the words damping, decay, and relaxation on an equal footing. However, strictly speaking, the isolated three-mode model cannot account for damping processes as in conventional open quantum system approaches~\cite{BreuerBook,WeissBook}, where information is irreversibly lost to the environment. Our quantum model experiences dephasing, similar to what is termed {\it  collapse and revival} of the wave function in quantum optics~\cite{1980EberlyPRL}. 
The collapse occurs since the initial state is a superposition  of exact Hamiltonian many body  eigenstates with eigenenergies that are anharmonic. This anharmonicity of many body eigenstates is the source of the ``collapse''-dephasing . Since the system has strictly speaking discrete spectrum, in addition to dephasing, an ``approximate revival'' of the intial state is expected to occur at the time scales of the order of the inverse of the smallest/typical  gap between neighboring levels in the energy spectrum (for the discussion in the context of thermalization see~\cite{Oliveira2018}). Still, the parallel of ``dephasing'' and thermalization in open systems can be drawn. Due to its complexity, our system plays the role of its own environment. Specifically,  the second excited mode can be understood as a minimal environment, while the system corresponds to the initially populated ground and first excited modes. This is an intuitive interpretation drawn from the fact that the second excited mode is not initially macroscopically occupied and its consideration represents the first step towards a more general approximation: one can consider that all Bogoliubov excitations are the environment for the two lowest modes in the same spirit as done when interpreting the Bose polaron motion as that of a quantum Brownian particle~\cite{Lampo2017}. 

We relate the decay of the oscillations with the fragmentation (loss of coherence) of the condensate and with the approach of the quantum three-mode model to the chaotic regime.  The connection between damping and fragmentation is in accordance with numerical studies done for a BEC in a quasi-1D bosonic Josephson junction using the multi-configurational time-dependent Hartree method for bosons (MCTDHB) \cite{Sakmann2014}. We also show that while the two-mode model can account for the loss of coherence and fragmentation, it does not show a quantum chaotic regime.   

Quantum chaos is associated with level repulsion and highly delocalized eigenstates, both of which guarantee the fast relaxation and thermalization of systems perturbed far from equilibrium~\cite{Gogolin2016,Borgonovi2016,Alessio2016}.  In the scenario of isolated quantum systems, fragmentation, relaxation, and thermalization are caused by the interparticle interaction, rather than by couplings with an external thermal bath. They therefore reinforce the fact that the mean-field approximation is not valid for long times.  

The paper is organized as follows. Section~\ref{sec:models} introduces the three-mode many-body model considered.  Sections~\ref{sec:dynamics} and \ref{sec:chaos} analyze how its properties change as the interaction strength increases from zero. In Sec.~\ref{sec:dynamics}, the dynamics under the two- and three-mode semiclassical models are compared with the quantum three-mode model. Two kinds of oscillations, their decay, and the phenomenon of fragmentation are discussed. Section~\ref{sec:chaos} addresses the onset of chaos and its connection with the decay of the oscillations. Conclusions are given in Sec.~V. We also present the study of the quantum dynamics for the two-mode model in App.~\ref{sec:two-mode} and discuss some previous results about the conditions for relaxation in isolated quantum systems in App.~\ref{sec:boundsRT}.

\section{Three-Mode Many-Body Model}
\label{sec:models}

The second-quantized Hamiltonian for $N$ ultracold bosons in an external potential $V({\bf x})$ is given by
\begin{equation}
\hspace{-1cm}H= \int d{\bf x}  \Psi^\dagger({\bf x}) \Big( -\frac{\hbar^2}{2m} \nabla + V({\bf x}) \Big) \Psi({\bf x})  + \frac{ g_{\rm{3D}}}{2} \int d{\bf x}  \Psi^\dagger({\bf x})\Psi^\dagger({\bf x})\Psi({\bf x}) \Psi({\bf x}),
\label{eq:Secondquant_Hamiltonian}
\end{equation}
where $\Psi({\bf x})$ represents the field operator, ${\bf x}\in\mathbb{R}^3$,  $\hbar$ is the Planck constant (which will be set to 1), $m$ is the mass of the particles, and $g_{\rm{3D}}$ is  the coupling constant in three dimensions governing the contact interactions. The hats on the operators are omitted to simplify the notation. Let us first consider that the trapping potential $V({\bf x})$  is parabolic in  the $x$ and $z$ directions, characterized by the trapping frequencies $\omega_{x,z}$. We assume the trapping in the $y$ direction is slightly anharmonic. In such case all single-particle eigenenergies are discrete and show accidental degeneracies given by $E_{j_x,j,j_z}=(1/2+j_x)\omega_x+(1/2+j_z)\omega_z+E_j$, with $j_{x,z}$ integers and $E_j$ the energy of $j$-th level of the separable single-particle Hamiltonian in the $y$ direction.  

For clarity of the explanation, we first introduce the one dimensional model. We assume that $\omega_{x,z}\gg E_{01}=E_1-E_0$, so that we can consider dynamics only in the transverse ($y$) direction. We aim for a model able to describe the dynamics of an initial state where a BEC is prepared in a coherent superposition of the lowest motional states along $y$.
We expand the field operator in terms of eigenfunctions of the single-particle Hamiltonian $\psi_j(\eta)$, $\eta=x,y,z$  i.e.  $\Psi=\sum_{j_x,j,j_z} a_{j_x,j,j_z} \psi_{j_x}(x)\psi_{j}(y)\psi_{j_z}(z)$, with $ a_{j_x,j,j_z}$ the annihilation operator. As the atoms are tightly confined in $x$ and $z$ the access to excited states in these directions is in practice forbidden, so the dynamics is frozen in those directions. Thus, as we only use in practice the operators $a_{0j0}$,  from here on we only keep the $j$ index. We note that there is a symmetry in the $y$ direction, as the Hamiltonian expressed in this basis is invariant by the reflection $y\to-y$.  According to our premises,  the atoms initially occupy macroscopically the two lowest modes in the $y$ direction. We truncate the expansion of the field operator in the third mode
\begin{equation}
\Psi = a_0 \psi_0 + a_1 \psi_1 + a_2 \psi_2
\end{equation}
 which produces the Hamiltonian 
\begin{eqnarray}
\hspace{-1cm}H_{\rm{3m}}& = \sum_i n_i E_i  + \sum_{i \neq j} U_{ij} \left[ (a_i^\dagger)^2 a_j^2 + (a_j^\dagger)^2 a_i^2 \right] + \sum_i U_i  n_i (n_i-1)  \nonumber \\
\hspace{-1cm}& + 4 \sum_{i \neq j} U_{ij} n_i n_j + 2 U_{1120} \left[ (a_1^\dagger)^2 a_0 a_2 + a_1^2 a_0^\dagger a_2^\dagger \right] + 4 U_{1120} \left[ n_1 a_0^\dagger a_2 + n_1 a_2^\dagger a_0 \right] \nonumber\\
\hspace{-1cm}&  + 2  U_{0222}  \left[ a_0^\dagger n_2 a_2 + a_2^\dagger a_0 n_2 \right]+ 2 U_{0002} \left[  a_2^\dagger n_0 a_0 + a_2 a_0^\dagger n_0 \right], 
\label{eq:MB3modes_Hamiltonian}
\end{eqnarray}
where  $ U_{ijkl} = (g/2)  \int\! dy\, \psi_i \psi_j \psi_k \psi_l$, $ U_{ij} = (g/2)  \int\! dy\, \psi_i^2 \psi_j^2$, $ U_{i} = (g/2)  \int\! dy\, \psi_i^4 $ (all wave functions are defined as real), and $E_i$ is the energy of level $i$.  Here, $g$ is the effective quasi-1D coupling constant along the $y$ direction. Figure~\ref{fig:3modes_pot} illustrates the parameters of the Hamiltonian~(\ref{eq:MB3modes_Hamiltonian}). The caption explains what each parameter denotes and the processes that the arrows represent.  We use the expansion in three modes because this is the minimal model that contains all important virtual process. For example, the term $a_0^\dagger a_2^\dagger(a_1)^2$ cannot be neglected when one assumes that the lowest two modes are macroscopically occupied,  $\langle a_{j}\rangle $ and $\langle a_{j}^\dagger\rangle $ are  $\sim \sqrt{N_j}$, $j=0,1$. In a BEC one assumes that fluctuations to other modes are negligible, and therefore neglects all quadratic terms with $j>0$. But in this case, the aforementioned term is not quadratic, thus giving a relevant contribution to the Hamiltonian. 
 In Appendix~\ref{sec:two-mode} we introduce the two-mode model, which we will compare with the results of the three-mode model obtained below. We  mention here that the  aforementioned reflection symmetry  $y\to-y$ translates into $\left[H_{\rm{3m}},P\right]$ with $ P= (-1)^{n_1}$, with $n_1=a^\dagger_1a_1$. As we discuss later,  identification of this symmetry is important in the study of quantum chaos below.      
 

%


To be able to model the experiments in~\cite{Bucker2013, vanFrank2014, vanFrank2016,Bonneau2018} we have to relax the one-dimensional assumption. In those experiments, a cigar-shaped BEC is produced in an elongated potential, i.e.  the potential is weakly confining along $x$ and more confining along $y$ and very tightly confined along $z$. It is still valid the assumption that the dynamics in the $z$ direction is frozen, as the excited states in that direction have very large energies. We assume that the wave function in the  longitudinal direction $x$ is well described with the Thomas-Fermi profile,  ${\rm TF} (x)$.
 In this paper we consider that initially part of the population is transferred to the first excited state. Particularly in all examples we assume that half of the population is initially excited. In such case, as we discuss later,  the period of the density oscillations that occur even in the non-interacting case is given by $E_{01}$ (see Sec.~\ref{sec:semiclass}). On the other hand the excitations along $x$ are much lower in energy, since $\omega_x<E_{01}$. These excitations can easily occur in the system, but they are much slower. For this reason we assume that along the evolution the system remains in  ${\rm TF} (x)$ along $x$ and study only the dynamics in $y$. With this, a  Hamiltonian formally identical to Eq.~(\ref{eq:MB3modes_Hamiltonian}) can be obtained, with a different expression of the coefficients. They have to be calculated taking into account that $N_0$ and $N_1$ correspond to the total population when integrating the corresponding excited mode in $y$ and ${\rm TF}(x)$, and thus all $ U_{ijkl}$ include the integration in $x$. This procedure gives rise to the interaction parameters gathered in Table~\ref{table:interactionpara}  for a system with $N=700$. We term these values $ U_{ijkl}^{\rm{exp}}$ as they are close to typical experimental values~\cite{Bucker2013, vanFrank2014, vanFrank2016, Bonneau2018}.  
We remark that, as  the trapping potential along $y$ is slightly anharmonic, the energy differences $E_{01}=E_1-E_0$ and $E_{12}=E_2-E_1$ are not equal.

In the numerical examples below, the interaction strengths are varied, so we take $ U_{ijkl}$ to be a constant $g$ multiplied by the  value from the table,  that is $U_{ijkl}=g\times U_{ijkl}^{\rm{exp}}$. Since the geometry of the trapping potential does not change, the orbitals $\psi_i$ do not change either. This means that the coupling constant $g^{\rm{num}}$ that we consider in the numerical examples is proportional to the experimental coupling constant, $g^{\rm{num}}=g\times g^{\rm{exp}}$. Then $g=1$ implies that $g^{\rm{num}}=g^{\rm{exp}}$.   

\begin{table}[h]
\begin{center}
\begin{tabular}{| l | l | l | l | l | l | l | l | l |}
\hline
$U_0$ & $U_1$ & $U_2$ & $U_{01}$ &  $U_{02}$ & $U_{12}$ & $U_{0112}$ & $U_{0002}$ & $U_{0222}$ \\
\hline
0.303 & 0.248 & 0.218 & 0.171 & 0.144 & 0.157 & -0.062 & 0.110 & -0.001 \\
\hline
\end{tabular}
\end{center}
\caption{Typical experimental values (in Hz) of the different parameters of the Hamiltonian in Eq.~(\ref{eq:MB3modes_Hamiltonian}). The energy differences between the levels are  $E_{01}=E_1-E_0= 1.770$kHz and $ E_{12}=E_2-E_1= 2.06$kHz. }
\label{table:interactionpara}
\end{table}

\section{Dynamics of the three-mode model}
\label{sec:dynamics}

With the Hamiltonian in Eq.~(\ref{eq:MB3modes_Hamiltonian}), we derive the equations of motion. In the Heisenberg picture, they are
\begin{equation}
i \frac{d a_j}{d t} = [a_j, H_{3m}], \hspace{1cm} j=0,1,2,
\end{equation}
which leads to
\begin{eqnarray}
i \frac{d a_0}{dt} & =  2 U_{01} a_0^\dagger a_1^2 + 2 U_{02} a_0^\dagger a_2^2 + 4 U_{01} n_1 a_0 + 4 U_{02} n_2 a_0 + 2 U_0 n_0 a_0 +  E_0 a_0\nonumber\\ 
&   + 2 U_{0112} ( a_2^\dagger a_1^2 + 2 n_1 a_2 ) + 2 U_{0222} n_2 a_2+ 2 U_{0002} (a_0^2 a_2^\dagger + 2 n_0 a_2 ), 
\label{eq:da0}
\end{eqnarray}
an equation similar to Eq.~(\ref{eq:da0}) for $a_2$, and
\begin{eqnarray}
i \frac{d a_1}{dt} & =  2 U_{01} a_1^\dagger a_0^2 + 2 U_{12} a_1^\dagger a_2^2 + 4 U_{01} n_0 a_1 + 4 U_{12} n_2 a_1   + 2 U_1 n_1 a_1 +  E_1 a_1\nonumber\\
&+ 4 U_{0112} ( a_0 a_2 a_1^\dagger + a_0^\dagger a_2 a_1 + a_2^\dagger a_0 a_1 ).
\end{eqnarray}

\subsection{Semiclassical dynamics}
\label{sec:semiclass}
To get physical insight into the role of the different parameters of Eq.~(\ref{eq:MB3modes_Hamiltonian}) and the processes represented in Fig.~\ref{fig:3modes_pot}, we consider a semiclassical version of the above equations of motion. For this,  we neglect quantum fluctuations and treat the operators as $c$-numbers, $\alpha_i = \sqrt{N_i} \exp(i \phi_i)$, where $N_i$ is the amplitude and $\phi_i$ is the phase of the fields $\alpha_i$ (for details on this procedure see~\cite{GarciaMarch2014}).  In this way, we obtain
\begin{eqnarray}
\frac{d N_0}{dt} &\!  =\!  - 4 U_{01} N_0 N_1 \sin 2 (\phi_0\! - \! \phi_1 )\!  -\!  4 U_{02} N_0 N_2 \sin 2 ( \phi_0\!  -\!  \phi_2 ) \\ 
& \!  + \! 4 U_{0112} N_1 \sqrt{N_0 N_2} [ \sin( 2 \phi_1 \! -\!  \phi_2 \! -\!  \phi_0 ) \! +\!  2 \sin( \phi_2 \! -\!  \phi_0 )]\nonumber \\
& \! + \! 4 U_{0222} N_2 \sqrt{N_0 N_2} \sin( \phi_2 \! -\!  \phi_0 ) \! + \! 4 U_{0002} N_0 \sqrt{N_0 N_2} \sin( \phi_2 \! -\!  \phi_0 ), \nonumber
\label{eq:N0_cfield}
\end{eqnarray}
\begin{eqnarray}
\frac{d \phi_0}{dt} & \! =\!  - 2 U_{01} N_1 \cos 2 (\phi_0 \! -\!  \phi_1 ) \! -\!  2 U_{02} N_2 \cos 2 ( \phi_0\!  -\!  \phi_2 ) \\ 
\hspace{-1cm} & \!  -\!  4  U_{01} N_1 \! -\!  4  U_{02} N_2 \! -\!  2  U_0 N_0 \! -\!  E_0 \nonumber\\
\! \! \! &\!  -\!  2 U_{0112} N_1 \sqrt{ N_2 / N_0} [ \cos( 2 \phi_1 \! -\!  \phi_2 \! -\!  \phi_0 ) \! +\!  2 \cos( \phi_2\!  -\!  \phi_0 )] \nonumber\\
\! \! \! & \! - \! 2 U_{0222} N_2 \sqrt{ N_2 / N_0} \cos( \phi_2\!  -\!  \phi_0 ) \!    \! -\!  6 U_{0002}  N_0 \sqrt{ N_2 / N_0} \cos( \phi_2\!  -\!  \phi_0 ), \nonumber
\label{eq:phi0_cfield}
\end{eqnarray}
with similar equations for $N_2$ and $\phi_2$, and
\begin{eqnarray}
\frac{d N_1}{dt}   &=  -  4 U_{01} N_0 N_1 \sin 2 (\phi_1 \! -\!  \phi_0 ) -  4 U_{12} N_1 N_2 \sin 2 ( \phi_1\!  -\!  \phi_2 )\nonumber\\
& -   8 U_{0112} N_1 \sqrt{N_0 N_2} \sin( 2 \phi_1 \! -\!  \phi_2 \! -\!  \phi_0 ),
\label{eq:N1_cfield}
\end{eqnarray}
\begin{eqnarray}
\frac{d \phi_1}{dt} & \! = \! -\!  2 U_{01} N_0 \cos 2 (\phi_1 \! -\!  \phi_0 ) \! -\!  2 U_{12} N_2 \cos 2 ( \phi_1 \! -\!  \phi_2 ) \\ 
& \!  -\!  4  U_{01} N_0 \! -\!  4  U_{12} N_2 \! -\!  2 U_1 N_1 \! -\!  E_1 \nonumber\\
& \! -\!  4 U_{0112} \sqrt{N_0 N_2} [\cos( 2 \phi_1 \! -\!  \phi_2 \! -\!  \phi_0 ) \! +\!  2 \cos( \phi_0 \! -\!  \phi_2 )].\nonumber
\label{eq:phi1_cfield}
\end{eqnarray}

\begin{figure}[htb]
\centering
	\includegraphics*[width=1.01\textwidth]{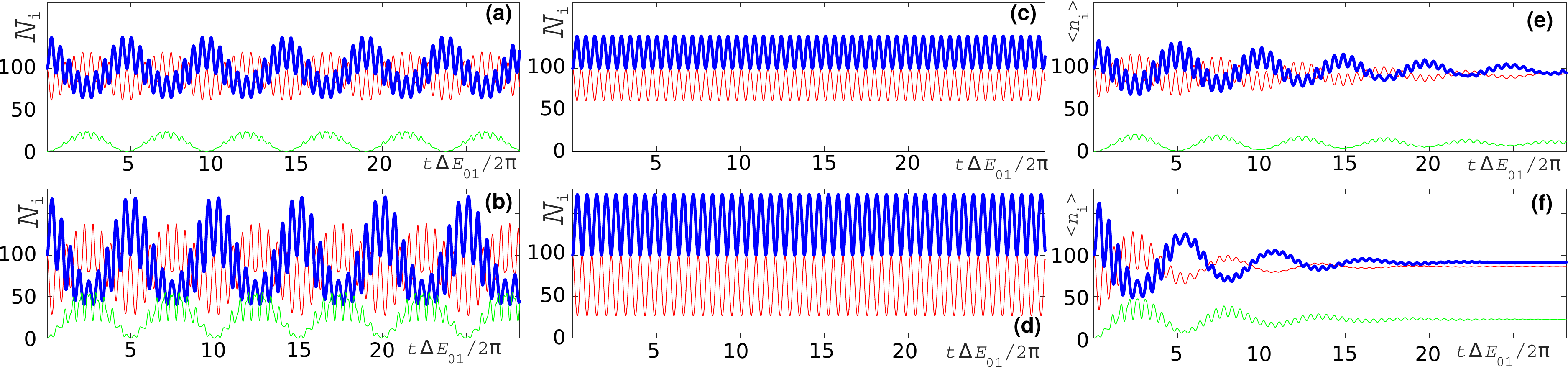}
	\caption{Evolution of the mode average occupations  for the semiclassical three-mode (left column), semiclassical two-mode (central column), and three-mode many-body (right column) models.   Upper (lower)  row corresponds to $g=10$ ($g=20$). We represent the ground mode with a red thin line, the first excited mode with a blue thick line, and the second excited mode  with a green thin line [the latter is the lower curves in panels (a), (b), (e) and (f)]. The initial condition is an equally weighted coherent superposition of the atoms in the ground and first excited states with a total number of atoms $N=200$. Time in all figures is adimensionalized with $\Delta E_{01}$ and divided by $2\pi$ to resemble approximately a period of oscillation.   }
	\label{fig2}
\end{figure}
\begin{figure}[htb]
\centering
	\includegraphics*[width=1.01\textwidth]{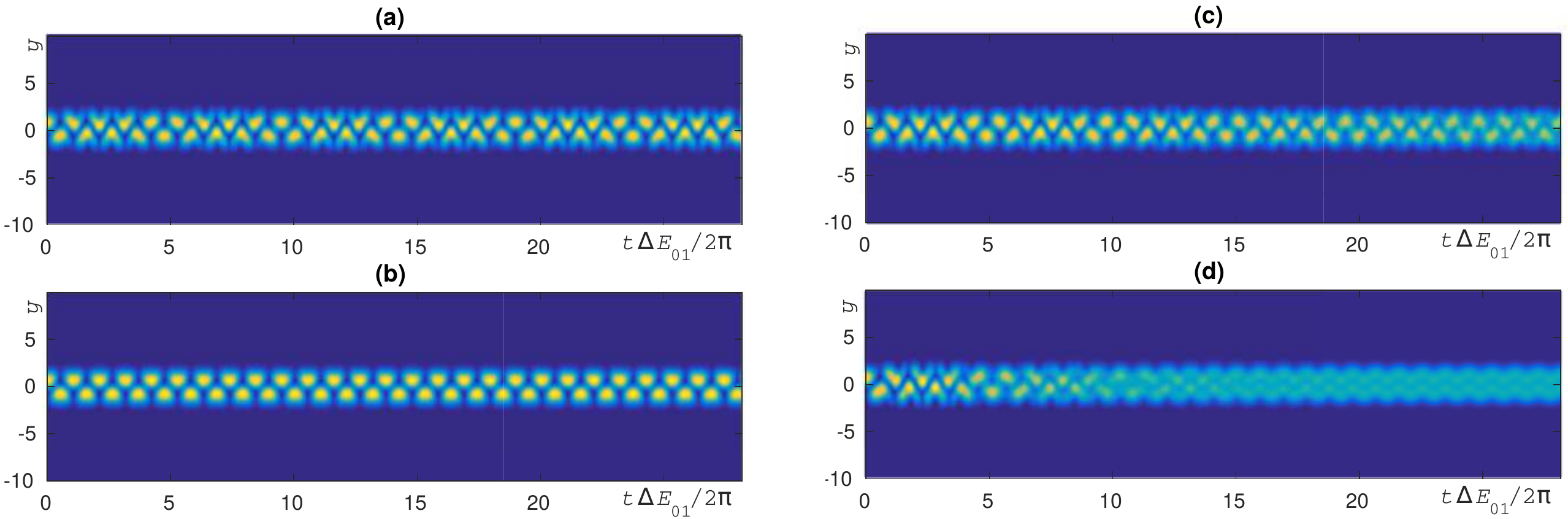}
 	\caption{Evolution of the density profiles for $N=200$ atoms (same initial state as in Fig.~\ref{fig2}). (a) Evolution of the density profile for the semiclassical three-mode model, $g=10$. (b)  Same for the semiclassical two-mode model. (c) and (d) Evolution of the density profile for the three-mode  many-body model for $g=10$ and $g=20$, respectively. Damping occurs in the many-body case after a number of oscillations. It occurs at a shorter time for larger interactions.  }
	\label{fig3}
\end{figure}

In Fig.~\ref{fig2} (a) and Fig.~\ref{fig2} (b) we show two exemplary dynamical evolutions of the amplitudes for the three modes. We assume that the initial condition is such that half of the atoms occupy the ground mode and the other half occupy the first excited mode. In particular, the initial state has $(N_0, N_1, N_2) = (100, 100, 0)$ atoms and all relative phases equal to zero. Two observations are made from the two panels. First, although the third mode is not populated initially, its gets significantly populated during the evolution, as we expected. Second, the dynamics presents two types of oscillations with different timescales. A fast oscillation with a period $T_\text{fast} \approx \SI{0.5}{\milli\second}$ and a slow oscillation with a longer period $T_\text{slow} \approx \SI{5}{\milli\second}$. 
 
From inspection of Eqs.~(\ref{eq:N0_cfield})-(\ref{eq:phi1_cfield}) one deduces that for $g=0$ (non-interacting limit, all $U$ coefficient vanish) the amplitudes $N_i$ are constant and the phases $\phi_i$ grow with a rate $E_i$. For the initial condition considered here  one observes density oscillations in the numerical simulations (not shown here) which are only due to these running phases. Thus, this is the main origin of the density oscillations observed in the density plots at finite $g$ (see Fig.~\ref{fig3}). Because of that we adimensionalize  the time in all figures with $\Delta E_{01}$.  For  $g$ small inspection of  Eq.~(\ref{eq:N0_cfield}) shows that $N_0$ will oscillate slightly around their initial value with amplitude proportional to $U_{01}$ and period proportional to $E_{01}$ [see first lines in Eqs.~(\ref{eq:N0_cfield}) and~(\ref{eq:N1_cfield})]. All other terms are small because $N_2$ is much smaller.  In our  numerical simulations with small $g$ (not shown) we observe that this is the only oscillation present in the populations. As $g$ is made larger, there is a part of population that occupies the second mode, so that $N_2$ is no longer negligible. In such case we observe a second type of oscillation, which is slower and has an amplitude  proportional to  $U_{0112}$.

For comparison, we show in Figs.~\ref{fig2} (c) and (d)  the amplitudes of the ground and first excited modes for a semiclassical two-mode model starting with the same initial state as in Figs.~\ref{fig2} (a) and (b). The evolution gets much simpler, as only the fast oscillation remains. The semiclassical equations for the two-mode model are trivially obtained by neglecting in Eq.~(\ref{eq:N0_cfield}-\ref{eq:phi1_cfield}) all variables and parameters associated with mode 2. They are equivalent to the equations of an oscillator as the bosonic Josephson junction~\cite{Gati2007}.  

In Fig.~\ref{fig3} (a) we depict the evolution of the corresponding density $|\psi(t,z)|^ 2$ for the example shown in Fig.~\ref{fig2} (a).  This evolution resembles qualitatively the initial density oscillations observed in the experiment and also reproduced with a quasi-1D GPE description of the dynamics along $y$~\cite{vanFrank2014,Bonneau2018}. On the other hand, the density evolution shown in Fig.~\ref{fig3} (b), which corresponds to the case in Fig.~\ref{fig2} (c), has no similarity with the experiment. These results confirm that the two-mode model is insufficient to describe even qualitatively the dynamics of the system and that the inclusion of the third mode is necessary to apprehend the complexity of the dynamics. As discussed in Sec.~\ref{sec:models}, the two-mode model offers an oversimplified picture of the system, as it neglects important processes that populate significantly the second mode. We checked that  involving more than three modes in the semiclassical description does not bring significant changes to the evolution.

The semiclassical three-mode model captures the initial oscillations as present also in the mean-field description. However, just as the GPE simulation, it does not describe any decay of these oscillations, as seen in Fig.~\ref{fig2} (a) and Fig.~\ref{fig2} (b) where the oscillations continue over time. 
In contrast, the many-body model discussed next accounts for a decay of the oscillations. We do not attempt here for a full study of the non-linear dynamics in the semiclassical equations, as our goals relies in the quantum dynamics discussed in what follows.

\subsection{Quantum many-body dynamics}
\label{sec:dyn}

To simulate the many-body dynamics, we perform the exact diagonalization of the many-body Hamiltonian in Eq.~(\ref{eq:MB3modes_Hamiltonian}). The dimension ${\cal D}$ of the Hamiltonian matrix is ${\cal D} = (N+m-1)!/[N! (m-1)!]$, where $m=3$ is the number of modes. For large $N$ ,  ${\cal D}\approx N^{m-1}$, as shown by using Stirling formula. 

The analysis of the system's time evolution via exact diagonalization is very general and can be adapted to different models, e.g. Lipkin-Meshkov-Glick~\cite{GarciaMarch2011,GarciaMarch2012,GarciaMarch2015} or Bose-Hubbard Hamiltonian~\cite{Zhang2010}. This approach has the advantage to be relatively simple. By comparison, other methods to describe many-body systems, such as the MCTDHB~\cite{Strelsov2006,Alon2008}, can include more features, but the physical phenomena at the origin of the observed features can be difficult to identify. Exact diagonalization is, however, limited by the exponential growth of the dimension ${\cal D}$ with each additional mode or particle. For the three-mode model, we simulate the evolution with a total number of atoms up to $N= 200$.

The system is initialized in a coherent superposition of ground and first excited states, which corresponds to the experimental initial state in~\cite{vanFrank2014,Bonneau2018},
\begin{equation}
| \psi_{\text{ini}} \rangle = \frac{1}{\sqrt{N!}} (c_0 a_0^{\dagger} + c_1 a_1^{\dagger})^N | \text{vac} \rangle
\label{Eq:InitialState}
\end{equation}
where  $ | \text{vac} \rangle$ is the vaccum and $ |c_0|^2 + |c_1|^2 = 1$. The initial average population of the two first modes is $\langle n_0 (0)\rangle = |c_0|^2 $ and $\langle n_1 (0)\rangle = |c_1|^2 $, with $\langle n_i(t) \rangle = \langle \psi(t) | a_i^{\dagger} a_i | \psi(t) \rangle$.  We present results for an initial coherent state with $c_0=c_1$, but no qualitative differences are observed when $c_{0} \neq c_{1}$. As expected, this initial state permits to reproduce faithfully the semiclassical results for small interacting systems. Even for largely interacting system, it reproduces the semiclassical results for the first few oscillations (see Fig.~\ref{fig2}). This is not the case if one takes a Fock initial state, i.e. $| \psi_{\text{ini}} \rangle = 1/(\sqrt{N_0!}\sqrt{N_1!}) (a_0^{\dagger})^{N_0}(a_1^{\dagger})^{N_1} | \text{vac} \rangle$ (generally we denote a Fock vector as $| \varphi_k\rangle = 1/(\sqrt{N_0!}\sqrt{N_1!}\sqrt{N_2!}) (a_0^{\dagger})^{N_0}(a_1^{\dagger})^{N_1}(a_2^{\dagger})^{N_2} | \text{vac} \rangle=\rangle = |N_0, N_1 ,N_2 \rangle$).  

To circumvent our limitation to relatively small system sizes, we increase the effective interaction constant to reach values of the product $g\,N$ that are close to those found experimentally~\cite{vanFrank2014,vanFrank2016,Bonneau2018}, where an example is given in table~\ref{table:interactionpara}, and hence  correspond to $g=1$ and $N=700$.   One needs to keep in mind, however, that keeping $g\,N$ constant, but varying the interaction parameter and atom number does not necessarily guarantee the same physical scenarios~\cite{Sakmann2014}.  For a fixed and small value of $g\,N$, a small interaction constant $g$ with a big value of $N$ ensures that the semiclassical description is valid, while a small number of atoms $N$ with large interactions leads to the strongly correlated quantum regime. To extend results obtained for low atom numbers (and high $g$) to the experimental case of high $N$ (and low $g$), a mapping to a known problem (e.g. a Bose-Hubbard model in a lattice) could provide some insight, but such mapping is not always trivial.

In Figs.~\ref{fig2} (e) and (f), we present two examples of the evolution of the occupation of the ground and first excited modes for $N=200$ for two values of $g$. In Figs.~\ref{fig3} (c) and (d), we show the corresponding density evolution. The figures make it evident that the many-body evolution differs from the semiclassical one. For the quantum model, the system damps to an equilibrium state after a few oscillations. The damping occurs earlier as $g N$ is made larger. We have not found any revival even for the longest numerical simulations performed (typically 15 oscillations in terms of $\Delta E_{01} /2\pi $, with the longest simulations up to $t_{\rm max}=30\,\Delta E_{01} /2\pi $).

\subsection{Damping of the oscillations}
\label{sec:damping}

The rest of the paper is devoted to investigating theoretically the origin of the damping of the  oscillations. We evaluate numerically the damping time $\tau$ as a function of $g \,N$. To this end, we simulate the system for a number of particles $N$ ranging between 40 and 200 and we also vary $g$. But before proceeding with this evaluation, it is beneficial to elaborate on some related points. 

\begin{figure}[htb]
\centering
	\includegraphics*[width=1.01\textwidth]{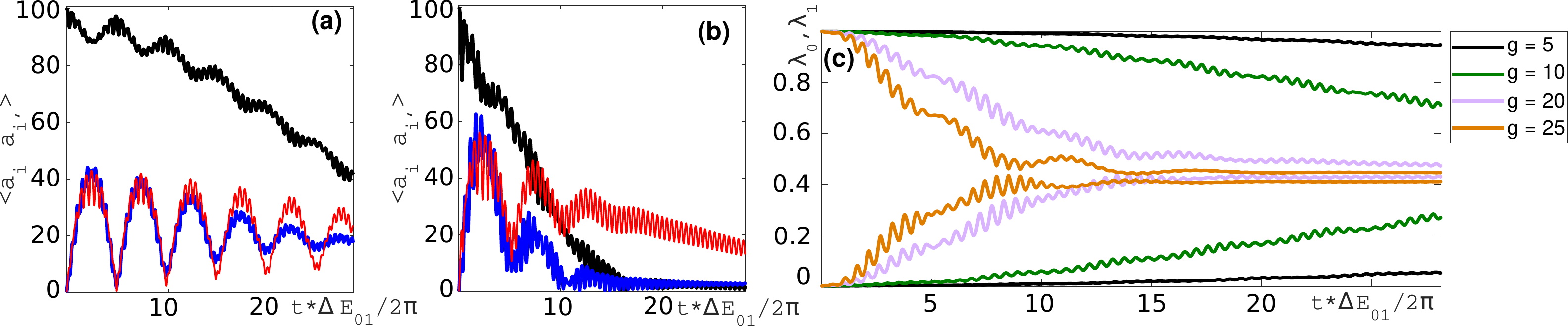}
 	\caption{Evolution of the off-diagonal elements of the one body density matrix (OBDM) $\langle a^\dagger_{i} a_{i'}\rangle$ when $g=10$ (a) and $g=20$ (b) [same initial state as in Fig.~\ref{fig2}].  Top black curve corresponds to $(i,i')=(1,2)$, red curve to $(i,i')=(1,3)$ and blue curve to $(i,i')=(2,3)$.  The damping observed in Fig.~\ref{fig2} (e), Fig.~\ref{fig2} (f), and Fig.~\ref{fig3} (d) is accompanied here by the vanishing of the off-diagonal correlations.  In (c): Evolution of the largest (top 4 curves) and second largest (bottom 4 curves) eigenvalues of the OBDM for different values of the interaction strength. At initial times, there is only one large eigenvalue, as expected for condensation in the coherent superposition of the ground and first excited mode. In time, the second largest eigenvalue becomes also sizeable, indicating fragmentation.  [The third eigenvalue, associated with the occupation of the second excited mode, is not shown, but it also becomes non-negligible.] This effect occurs at shorter times for larger interaction strengths. Thus, the damping in Fig.~\ref{fig2} (e), Fig.~\ref{fig2} (f), and Fig.~\ref{fig3} (d), the loss of coherence in panels (a) and (b), and the fragmentation in (c) occur together. 	}
	\label{fig4}
\end{figure}

First, we note that the decay of the oscillations of all modes occupations, $\langle a_i^\dagger a_i\rangle$ for $i=0,1,2$, are accompanied by a decay to zero of the coherence terms $\langle a_i^\dagger a_j\rangle$ with $i\ne j$. This parallel is verified by comparing the decay of the oscillations in Fig.~\ref{fig2} (e), Fig.~\ref{fig2} (f), and Fig.~\ref{fig3} (d) with Figs.~\ref{fig4} (a) and (b). The vanishing of the off-diagonal terms is associated with the fragmentation of the BEC, as discussed next. 

The phenomenon of fragmentation can be understood as follows. 
For a BEC in a trap, when there is only one large eigenvalue of the one-body density matrix (OBDM), $\rho=|\Psi\rangle\langle \Psi |$, the semiclassical Gross-Pitaevskii approach is appropriate. In this case, the depletion cloud of atoms which are not occupying the condensate is small.  In contrast, the presence of more than one large eigenvalue indicates that the depletion cloud is large and that many atoms are not Bose-Einstein condensed~\cite{PethickBook}. For instance, in the context of double-well potentials (and generally with two-mode models),  when there are two, and only two, large eigenvalues, the system is said to be fragmented~\cite{Spekkens1999,Mahmud2005}. This means that the atoms occupying the two distinct modes of the system cease to be coherent, while coherence may still exist among the atoms occupying each individual mode.   In general, fragmentation corresponds to the separation of an initially fully condensed state into two or more independent condensed parts. This scenario can be pictured as a double-well potential with an infinitely large barrier between the wells, so that the system is effectively cut in two halves, with no coherence among them. Of course, a single-shot experiment can show fringes and interference between the two condensates (see discussion in pg. 343  of~\cite{Leggett2001}, where interference experiments in the Fock regime in a double well is discussed).  We clarify here that it is in this sense that we talk about loss of coherence. 

In Fig.~\ref{fig4} (c), we show the time evolution of the two largest eigenvalues of the OBDM for different values of $g N$, taking $N=200$.  Initially, there is only one large eigenvalue. It decreases in time, while the second largest eigenvalue increases. After the damping occurs, one finds three eigenvalues that are significantly different from zero [only two are shown in Fig.~\ref{fig4} (c), but the third one is also non-negligible, as the sum of the first two does not amount to 1]. This indicates fragmentation in the three modes. 

The loss of coherence, just as damping, occurs earlier in time as $g N$ is made larger [compare Fig.~\ref{fig2} (e), Fig.~\ref{fig2} (f), and Fig.~\ref{fig4} (c)]. After relaxation, the system is found in a Fock state, that is, a state with a determined value of atoms occupying each mode. The link between damping and fragmentation has also been pointed out in two-mode models~\cite{Sakmann2009,Sakmann2014}.  

\begin{figure}[htb]
\centering
	\includegraphics*[width=0.5\textwidth]{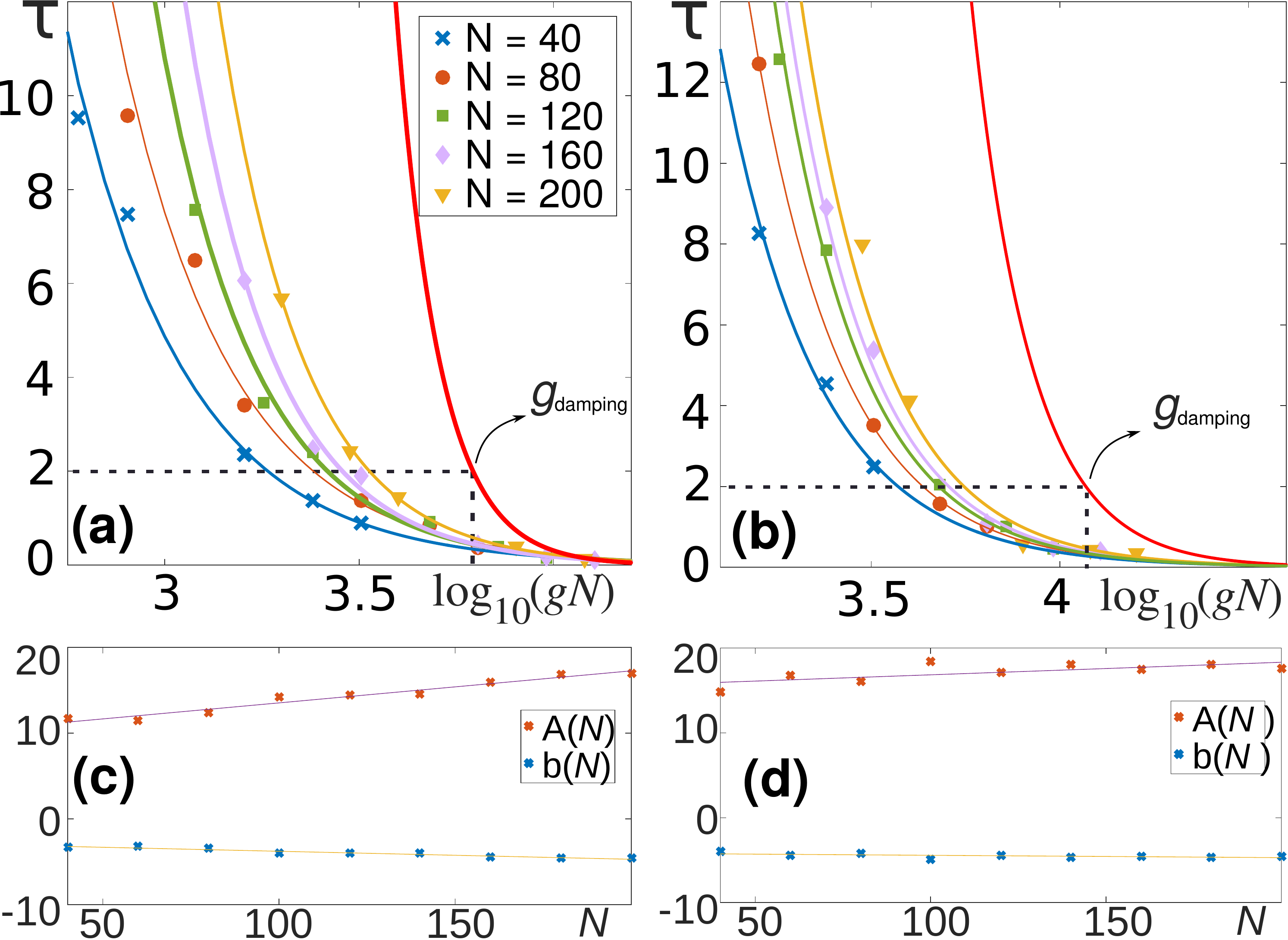}
 	\caption{Damping  time $\tau$ as a function of $g\,N$ (top panels).  The damping time is obtained from the many-body simulations for atom numbers from $N=40$  to $N=200$. The damping time is defined as the time at which the largest eigenvalue of the OBDM is smaller than 0.98 [panel (a)] or 0.85 [panel (b)].  The lines correspond to fitting curves of the form  $\tau = a(N) \exp \left[b(N) x \right]$, $x=\log_{10}(g N)$,  to the numerical results. We also show the resulting parameters $A(N)=\ln[a(N)]$ and $b(N)$ in panels (c) and (d).   The red thick line in (a) and (b) shows the expected behavior for $N=700$ obtained by extrapolating the numerical interpolated curves shown in (c) and (d). In panels (a) and (b) and for $N=700$,  we also represent $g_{\rm{damping}}$ , defined as the value of the coupling constant for which $\tau=2$. }
	\label{fig5}
\end{figure}

We estimate the damping time $\tau$ from the evolution of the eigenvalues of the OBDM.  In Fig.~\ref{fig5}, we plot our numerical estimates for $\tau$ as a function of $g\,  N$.  Two numerical criteria are used to determine the damping time. In Fig.~\ref{fig5} (a), $\tau$ is the time at which the largest eigenvalue of the OBDM gets smaller than 0.98. In Fig.~\ref{fig5}  (b), $\tau$ is the time when the largest eigenvalue becomes smaller than 0.85. 

For each $N$, we observe that the dependence of $\tau$ on $g N$ can be fitted to a  function of the form $\tau = a(N) \exp \left[b(N) x \right]$, with $ x=\log_{10}(g N)$.  For every value of  $N$, we fit the parameters $ a(N)$ and $b(N)$ from the results of the numerical simulations performed with a large number of values of $gN$. We present the results for $A(N)=\ln[a(N)]$ and $(b(N)$ in Fig.~\ref{fig5} (c) and (d). The corresponding fitted curves are the solid lines in Fig.~\ref{fig5} (a) and (b). 
For every curve corresponding to a different $N$, we define  $g_{\rm{damping}}$ as that in which the damping time is reduced to two oscillations in terms of $\Delta E_{01} /2\pi $ [see Fig.~\ref{fig5} (a) and (b)]. With the time adimensionalization we used, this corresponds to $\tau=2$. We use two oscillations as a criteria to define  $g_{\rm{damping}}$ because we observe that, in this  way, for $g<g_{\rm{damping}}$ the damping time increases significantly as one decreases $g$. This allows to distinguish from the region with $g>g_{\rm{damping}}$, where the damping time is reduced  strongly.  For the large values of $g$ and the atom numbers considered in Fig.~\ref{fig5},  all the curves for different $N$ show very similar behaviors.

Our numerical studies are limited to numbers of atoms up to $N=200$, but one can extrapolate the damping times to the case with $g=1$ and $N=700$, which is the lowest number of atoms performed in the experiment (see our convention as described in the end of Sec.~\ref{sec:models}).  To this end, we fit a straight line to the coefficients $a(N)$ and $b(N)$ [see fitting lines in  Figs.~\ref{fig5} (c) and (d)]. We use these fitted behavior to extrapolate the behavior to larger number of atoms to estimate the coefficients $a(700)$ and $b(700)$  which correspond to the curve expected for $N=700$. The result is the thick red curve in  Figs.~\ref{fig5} (a) and (b), where the estimated coefficient $A(700)\approx 35$ and 25, respectively, while $b(700)\approx -9$ and -6, respectively. 
According to these fittings,  the damping occurs around $\log_{10}(g N)=3.6$ $[3.8]$ with $N=700$ for the criterion used in Fig.~\ref{fig5} (a) [(b)]. This corresponds to $g_{\rm{damping}}\approx 5$ [$g_{\rm{damping}}\approx 9$],   while in the experiment it happens at $g_{\rm{damping}}=1$ (following our convention).  The damping time predicted for $g=1$ and $N=700$ from the curves of Figs.~\ref{fig5} (a) and (b) is of the order of thousands of oscillations, much larger than the one observed in the experiment ($\tau_{\rm{exp}}\,\Delta E_{01}\le 15$) \cite{vanFrank2014,Bonneau2018}.  Thus, even though the three-mode quantum model describes a damping of the density oscillations similarly to the experimental observations~\cite{Bonneau2018}, the damping timescale it predicts differs from the experimental one. 
 Some other effects may cause the shorter damping time seen in the experiment, such as dephasing dynamics in the longitudinal direction, perpendicular to the one-dimensional plane that we consider here. 

In the next section, we explore the relationship between the onset of quantum chaos, the decay of the oscillations, and the fragmentation of the condensate. We numerically link the finite values of $\tau$ with the approach of the quantum model to the chaotic regime.


\section{Onset of Quantum Chaos and Damping}
\label{sec:chaos}

Isolated many-body quantum systems perturbed far from equilibrium relax quickly to a new equilibrium despite the absence of external couplings. The driving mechanism for the equilibration is the internal couplings between particles~\cite{Borgonovi2016}. While both integrable and chaotic systems undergo a similar process, relaxation to thermal equilibrium is expected only for chaotic systems.

The many-body three-mode model undergoes a transition from integrability to chaos as the interaction strength $g$ increases from zero. This is in contrast with the many-body two-mode model, which is integrable for any value of the interaction, as discussed in App.~\ref{sec:two-mode}.

In the many-body three-mode model, as the number of atoms $N$ increases, smaller values of $g$ are needed to move the system away from the integrable limit. This behavior mirrors our findings for the damping time, which also decreases as $gN$ gets larger.

\subsection{Quantum Chaos}
Quantum chaos refers to signatures observed at the quantum level that indicate that the classical counterpart of the system is chaotic. The concept has been extended to any quantum system that exhibits those properties even if it does not have a classical limit. A main signature of quantum chaos is level repulsion and the consequent rigidity of the spectrum. 

\subsubsection{Level Spacing Distribution}
There are different ways to detect level repulsion and therefore the crossover from integrability to quantum chaos~\cite{Guhr1998}. The most commonly used quantity is the distribution $P$ of the spacings $s$ between neighboring unfolded levels. In integrable models, the levels can cross and the distribution is usually Poisson,
\[
P_{ P}(s) = \exp(-s),
\]
although variations may be found. This may occur, for example, in systems with an excessive number of degeneracies or in ``picket-fence'' spectra where the eigenvalues are nearly equally spaced. A typical example for the latter is the case of uncoupled harmonic oscillators~\cite{Berry1977b,Pandey1991}.
In chaotic systems, on the other hand, crossings are avoided and $P(s)$ follows the Wigner-Dyson distribution, as predicted by random matrix theory. In systems described by Hamiltonian matrices that are real and symmetric, the shape of this distribution is given by
\[
P_{ WD}(s) = \frac{\pi s}{2}\exp \left( -\frac{\pi s^2}{4}\right).
\]

\begin{figure}[htb]
\centering
\includegraphics*[width=0.8\textwidth]{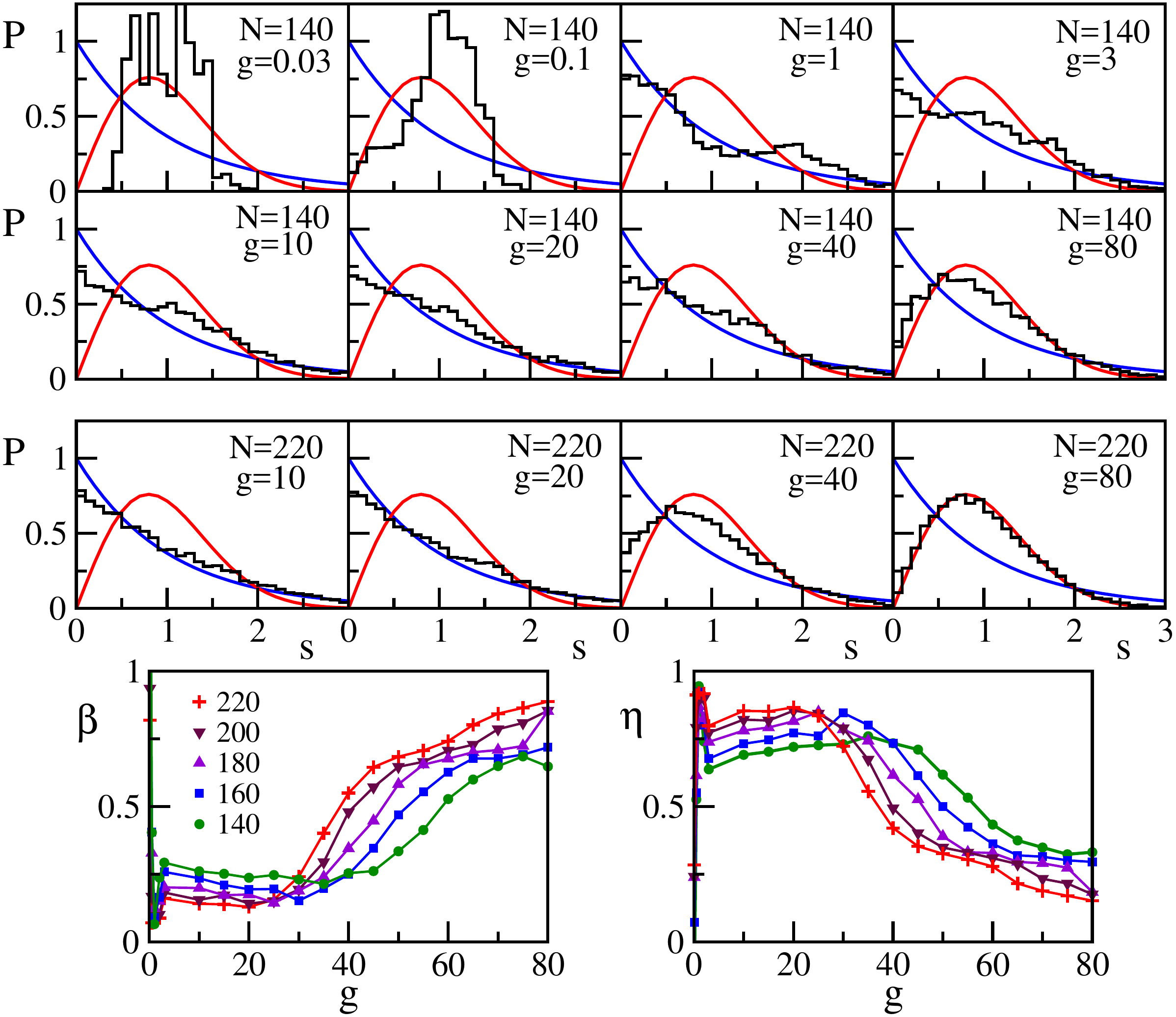}
\caption{Level spacing distribution (three top rows) for the three-mode model for two numbers of particles and different values of  the interaction strength, as indicated in the panels. Curves for the Poisson and Wigner-Dyson distributions are also presented for comparison. The two bottom panels show chaos indicators $\beta$ and $\eta$ as a function of the interaction strength for different $N$'s. The results for all panels are averaged over the two parity sectors.  Note that  $\beta\to1$ and $\eta\to0$ implies quantum chaos (approach to a Wigner-Dyson distribution). }
\label{fig:WD}
\end{figure}

In Fig.~\ref{fig:WD}, we compare the level spacing distribution of the three-mode quantum model for different values of $g$ and $N$. To get a meaningful distribution, the levels need to be separated by symmetry sector~\cite{Santos2009JMP}. Following the description of Eq.~(\ref{eq:MB3modes_Hamiltonian}), we separate the eigenvalues by the parity of the eigenstates. The two top rows of Fig.~\ref{fig:WD} are obtained for $N=140$. When the interaction strength is small, $g<10$, and the model is close to integrability, the level spacing distribution is not even Poisson. The distributions for $g=0.03, 0.1$ suggest a ``picket-fence'' spectrum~\cite{Lewenkopf1991}. For large interaction, $g>40$, the transition to the Wigner-Dyson distribution is clear. 

The second and third rows of Fig.~\ref{fig:WD} compare $P(s)$ for two different choices of $N$. As the number of atoms increases, the transition to chaos occurs for smaller values of the interaction~\cite{Santos2010PRE}. This is evident by contrasting the panels with strong interactions ($g=40$ and $g=80$) for $N=140$ with those for $N=220$. This indicates that when $N$ is very large, infinitesimal interactions may suffice for the onset of quantum chaos. 

The two bottom panels of Fig.~\ref{fig:WD}  show results for chaos indicators $\beta$ and $\eta$. These are measures of the proximity of $P(s)$ to Poisson or to Wigner-Dyson distributions. The indicator $\beta$ is obtained by fitting $P(s)$ with the Brody distribution~\cite{Brody1981},
\[
P_B(s) = (\beta +1) b s^{\beta} \exp \left( -b s^{\beta +1} \right), 
\hspace{0.2cm}
b= \left[\Gamma \left( \frac{\beta + 2}{\beta +1} \right)\right]^{\beta +1},
\]
where $\Gamma$ is the Euler's gamma function.
When $\beta=0$, the distribution is Poisson and for $\beta=1$, $P(s)$ has the Wigner-Dyson shape. The indicator $\eta$ was introduced in \cite{Jacquod1997} and is defined as
\begin{equation}
\eta = \frac{\int_0^{s_0} [P(s) - P_{WD}(s)] ds}{\int_0^{s_0} [P_P(s) - P_{WD}(s)] ds}, 
\end{equation}
where $s_0$ is the first intersection point of $P_P(s)$ and $P_{WD}(s)$. For a Poisson distribution, $\eta \rightarrow 1$, and for the Wigner-Dyson distribution, $\eta \rightarrow 0$.

In Fig.~\ref{fig:WD}, the results for $\beta$ and $\eta$ for $g<10$ need to be taken with care. For the numbers of atoms accessible to us, this range of interaction strengths leads to shapes other than $P_P$, $P_{WD}$, or any intermediate distribution between the two, as seen in the first row of Fig.~\ref{fig:WD}. The fact that for $g<10$, the indicator $\beta$ ($\eta$) increases (decreases) as $g$ and $N$ get smaller simply indicates that we move away from the Poisson distribution, but this is not accompanied by an approach to $P_{WD}$. We instead approach the integrable point of three uncoupled oscillators, $H \sim \sum_i n_i E_i$. 

For the numbers of atoms considered here, the transition from Poisson to Wigner-Dyson is well captured by the chaos indicators when $g>10$. The plots for $\beta$ and $\eta$ in Fig.~\ref{fig:WD} reinforce our statement above that the transition to chaos happens for smaller values of $g$ as $N$ increases.

\subsubsection{Structure of the Eigenstates}
The emergence of random matrix statistics is tightly connected with the appearance of chaotic eigenstates, that is states that are highly delocalized and fill the energy shell~\cite{Flambaum1997,Santos2012PRL,Santos2012PRE}. To measure the level of delocalization of the eigenstates $| \psi_{\nu} \rangle$, one can use quantities such as the participation ratio,
\begin{equation}
\text{PR}^{(\nu)}=\frac{1}{\sum_j | C_j^{(\nu)}|^4},
\label{eq:PR}
\end{equation}
where $C_j^{(\nu)} = \langle \varphi_j | \psi_{\nu} \rangle$ is the overlap between the eigenstate $|\psi_{\nu}\rangle$ and the basis vector $ |\varphi_j \rangle$.  PR is large when the eigenstate is delocalized in the chosen basis. The choice of basis for the analysis of the structure of the eigenstates is physically motivated. For the three-mode model, we select the Fock basis, $ |\varphi_j \rangle = |N_0, N_1 ,N_2 \rangle$. In the absence of interaction, when the eigenstates coincide with the basis vectors, PR=1. 

\begin{figure}[htb]
\centering
\includegraphics*[width=0.8\textwidth]{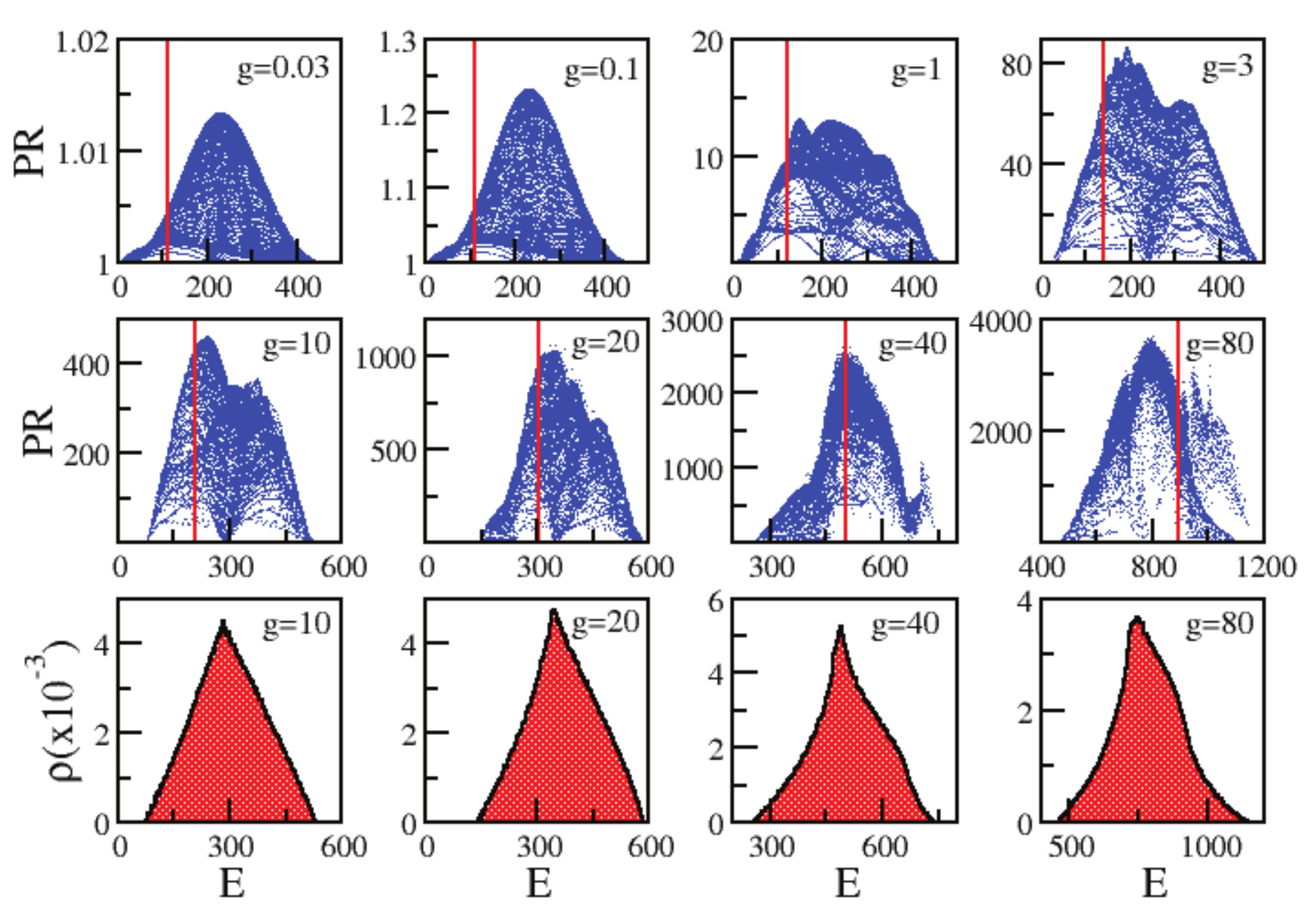}
\caption{Participation ratio (two top rows) and density of states $\rho$ (bottom row) for the three-mode model with different values of the interaction strength (indicated); $N=220$. Both parity sectors are included. Vertical lines mark the energy of the initial state.}
\label{fig:PR3}
\end{figure}

Each panel of the two top rows of Fig.~\ref{fig:PR3} show the values of PR for all eigenstates. Different values of $g$ are considered. The level of delocalization increases significantly with the interaction strength. One also notices that the highest values of PR occur close to the middle of the spectrum. This reflects the shape of the density of states $\rho$, shown in the bottom row of Fig.~\ref{fig:PR3} for comparison~\cite{footnote}. The density of states peaks close to the middle of the spectrum, where the largest concentration of eigenstates is found. This is the region where we expect the eigenstates to be more delocalized states, while at the borders, PR is smaller. 
 
The middle row of Fig.~\ref{fig:PR3}  illustrates the consequence of the transition to chaos. For $g<40$ and thus away from the chaotic regime, there are large fluctuations in the values of PR. This implies that eigenstates very close in energy can have very different levels of delocalization. In contrast, in the chaotic region ($g=80$), the structures of the eigenstates become very similar, especially close to the middle of the spectrum, where PR becomes a smoother function of energy. At this point, the states approach random vectors. The similarity between eigenstates very close in energy is what guarantees the validity of the eigenstate thermalization hypothesis (ETH) and the viability of thermalization~\cite{Santos2010PRE,RigolSantos2010,Santos2010PREb}, as discussed next.

\subsection{Thermalization}
\label{sec:thermalization}

The analysis of the onset of thermalization involves two steps. First one needs to ensure that the system equilibrates. Next, we verify whether the equilibrium is thermal or not. 

\subsubsection{Equilibration}
How the isolated system reaches equilibrium is the subject of the broad field of nonequilibrium quantum dynamics to which the previous section and several other works have been devoted to, including studies about pre-thermalization~\cite{Gring2012,MoriARXIV}. A brief discussion about the subject is presented in App.~\ref{sec:boundsRT}. In this section, we are concerned with the equilibrium point itself.  

One can say that isolated quantum many-body systems without too many degeneracies equilibrate, because revivals become rare and take exceedingly long times to happen as the system size increases. For all practical purposes, the coherences are irreversibly lost. The systems equilibrate in a probabilistic sense. To better explain what we mean by this, consider a general observable $O$ evolving in time according to the equation
\begin{eqnarray}
O(t) &= \langle \psi(0) | e^{i H t} O e^{-i H t} | \psi(0) \rangle\nonumber \\
&=  \sum_{\nu \neq \mu=1}^{ {\cal D} } C_{\text{ini}}^{\mu *} C_{\text{ini}}^{\nu} O_{\mu \nu} e^{i (E_{\mu} - E_{\nu}) t} + \sum_{\nu=1}^{ {\cal D} } |C_{\text{ini}}^{\nu}|^2 O_{\nu \nu} ,
\label{Otime}
\end{eqnarray}
where $ | \psi_{\nu} \rangle$ and $E_{\nu}$ are the eigenstates and eigenvalues of $H$, ``$\text{ini}$'' indicates the initial state, $O_{\mu \nu}  = \langle \psi_{\mu} | O | \psi_{\nu} \rangle$, and $O_{\nu \nu} $ is the eigenstate expectation value (EEV) of $O$. After a transient time, the system is said to have reached a new equilibrium if $O(t)$ simply fluctuates around the infinite-time average, 
\begin{equation}
\overline{O} = O_{\text{DE}} =\sum_{\nu=1}^{ {\cal D} } |C_{\text{ini}}^{\nu}|^2 O_{\nu \nu} ,
\label{OtimeAve}
\end{equation}
and remains very close to this value for most times. Since the infinite-time average only involves the diagonal matrix elements $O_{\nu \nu} $, this average is often referred to as ``diagonal ensemble'' (DE) average. 

To talk about equilibration, it is therefore essential that the fluctuations around $O_{\text{DE}}$ be small and decrease with system size~\cite{Srednicki1996,Srednicki1999,Reimann2008,Linden2009,Linden2010,Short2011,Short2012,Reimann2012,Venuti2013,Zangara2013,Kiendl2017}. 
Equilibration does not require chaos in the sense of level repulsion, but it needs highly delocalized eigenstates, delocalized initial states, and not too many degeneracies.

 As shown in Figs.~\ref{fig2} (e) and (f), the three-mode model relaxes to a Fock state with a fixed number of atoms occupying each mode, that is $\langle \psi(t) | n_{0,1,2} | \psi(t) \rangle$ decays to $\overline{n}_{0,1,2}$.  The fluctuations after equilibration are small and decrease with $g$, as one sees by comparing Fig.~\ref{fig2} (e) and Fig.~\ref{fig2} (f). 

\subsubsection{Thermal Equilibrium}
 
After equilibration, the observable will have reached thermal equilibrium if its infinite-time average coincides with a thermodynamic average, that is if
\begin{equation}
O_{\text{DE}} = O_{\text{ME}} ,
\label{Eq:thermal}
\end{equation}
where 
\begin{equation}
O_{\text{ME}} \equiv  \frac{1}{  {\cal{N}}_{E_{\text{ini}},\delta  E}  }  \sum_{\begin{array}{c}\nu \\ |E_{\text{ini}}-E_\nu|<\delta E\end{array}} \hspace{-0.5cm} O_{\nu \nu} 
\end{equation}
is the average over a microcanonical ensemble and  ${\cal{N}}_{E_{\text{ini}},\delta E}$ is the number of energy eigenbases in the window $\delta E$ taken around the energy $E_{\text{ini}}$ of the initial state. Equation~(\ref{Eq:thermal}) holds when $O_{\nu \nu} $ for eigenstates close in energy coincide with the microcanonical average, an idea that is at the heart of statistical mechanics and has become known as eigenstate thermalization hypothesis (ETH). 

When studying thermalization in finite systems, we investigate how close the left and right sides of Eq.~(\ref{Eq:thermal}) are and whether they approach each other as the system size increases. This is guaranteed to happen when the eigenstates are nearly random vectors. All random vectors are equivalent, since their components are simply random numbers. Thus,  $O_{\nu \nu} $ computed with one random vector is very similar to the result for any other random vector, apart from small fluctuations that decrease with system size. 

\begin{figure}[htb]
\centering
\includegraphics*[width=0.8\textwidth]{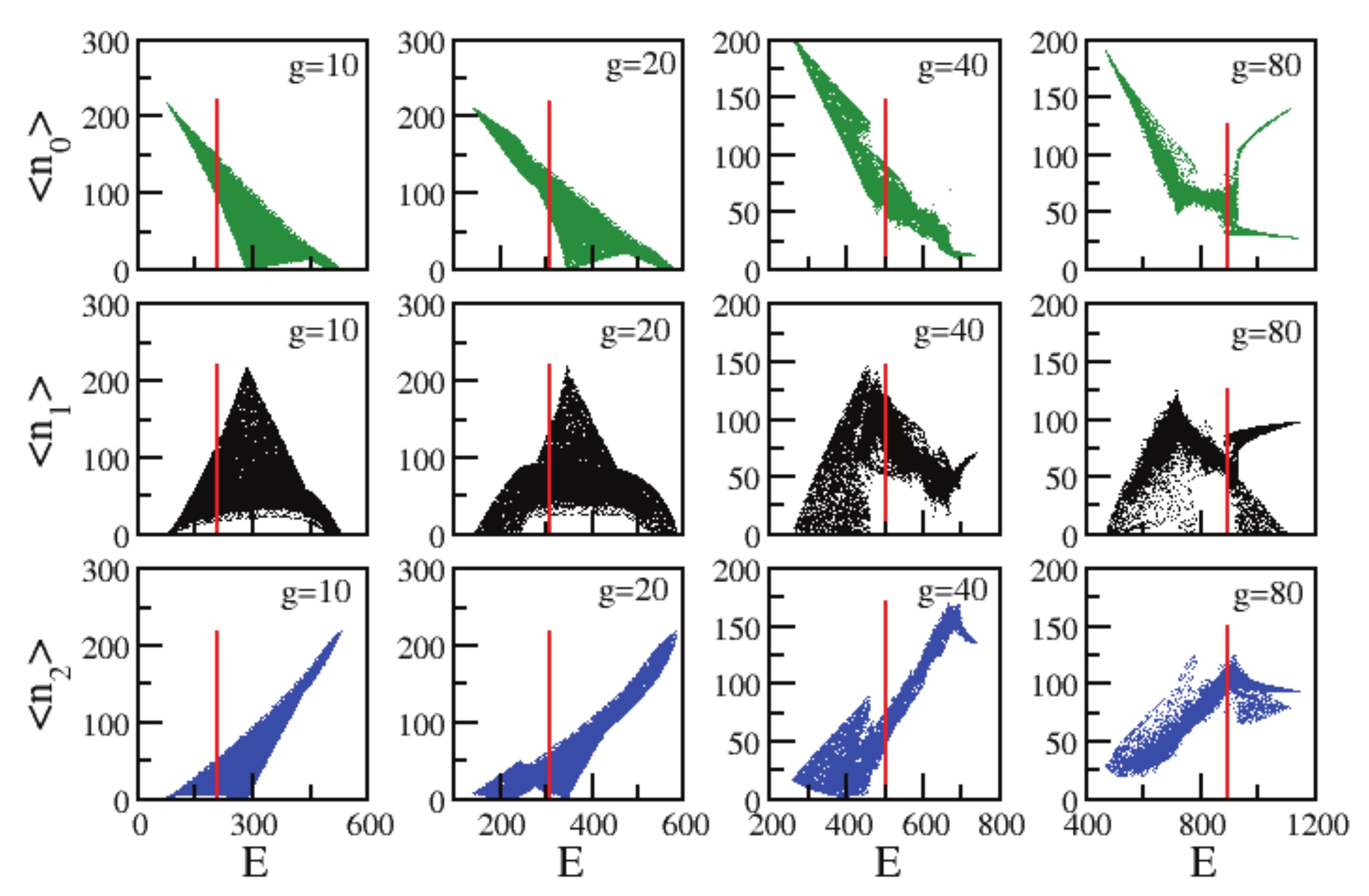}
\caption{Eigenstate expectation value (EEV) for $\langle n_{i}\rangle$, $i=0,1,2$, for all eigenstates (both parity sectors are included) and different values of the interaction strength (indicated); $N=220$. Vertical lines mark the energy of the initial state. }
\label{fig:EEV}
\end{figure}

In full random matrices, all eigenstates are random vectors, in which case thermalization is trivial. In realistic systems, the eigenstates away from the borders of the spectrum approach random vectors as the system moves toward the chaotic regime (see the discussion about the results for PR when $g=80$ in Fig.~\ref{fig:PR3}). This is paralleled by the behavior of the EEV for the occupations of the three modes depicted in Fig.~\ref{fig:EEV}. As $g$ increases, the fluctuations decrease and the EEVs show a smoother behavior with energy, especially close to the middle of the spectrum.

\begin{figure}[htb]
\centering
\includegraphics*[width=0.8\textwidth]{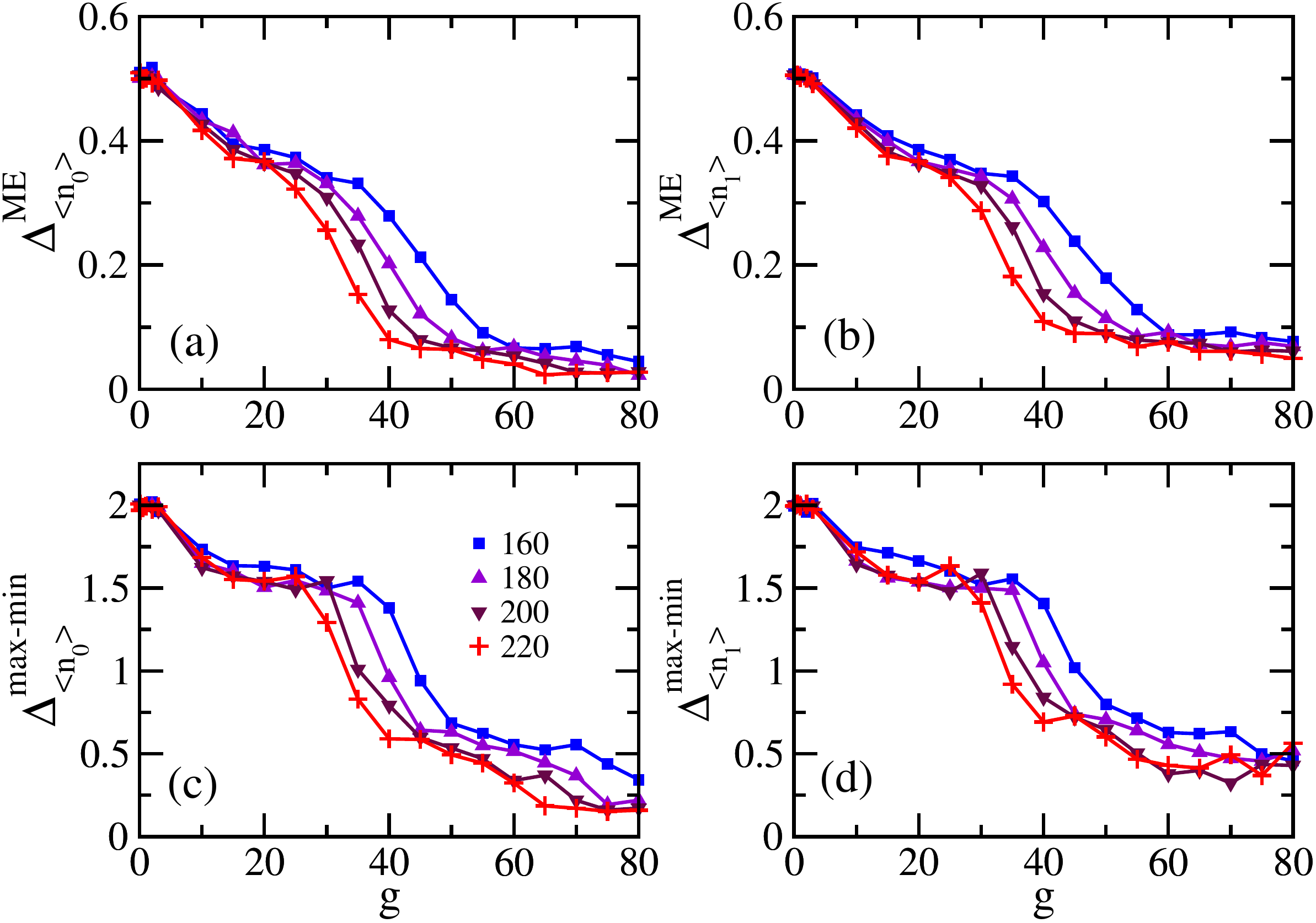}
\caption{Relative difference between the eigenstate expectation value (EEV) and the microcanonical ensemble (ME) [(a) and (b)], and normalized extremal fluctuations of EEV [(c) and (d)] as a function of $g$ and for different numbers $N$ of atoms (indicated). All eigenstates of both parity sectors in the window $[-\delta E, \delta E]$ with $\delta E=0.5$ are taken into account.}
\label{fig:Rel}
\end{figure}

To quantify the proximity of the EEV to the microcanonical average, we compute~\cite{RigolSantos2010}
\begin{equation}
\Delta^{\text{ME}}_{O} = \frac{  \sum_{\nu} |O_{\nu \nu} -O_{\text{ME}} |  }{  \sum_{\nu} O_{\nu \nu} } ,
\end{equation}
where for the three-mode model, $O=\langle n_0 \rangle, \langle n_1 \rangle,\langle n_2 \rangle$. The sum includes only the eigenstates within the microcanonical  window $[E-\delta E, E+\delta E]$. In Fig.~\ref{fig:Rel}, we choose $E$ very close to the middle of the spectrum and $\delta E=0.5$, so that the microcanonical window contains approximately $10^2$ levels. Provided there is a reasonable number of levels inside the window, the precise value of $\delta E$ does not affect the results. Similarly to what we find for the chaos indicators in Fig.~\ref{fig:WD}, $\Delta^{\text{ME}}_{O}$ in Figs.~\ref{fig:Rel} (a) and (b) decreases with $g$ and also with $N$, suggesting that the fluctuations vanish in the thermodynamic limit.

A more stringent demonstration of the vanishing of the fluctuations for strong interactions and large numbers of particles is made with the normalized extremal fluctuation of $O$, defined as~\cite{Santos2010PREb},
\begin{equation}
\Delta^{\text{max-min}}_{O} = \left| \frac{  \max{O} - \min{O}  }{  O_{\text{ME}} } \right|.
\end{equation}
The maximum ($\max{O}$) and minimum ($\min{O}$) values of the EEV are obtained for the eigenstates within the microcanonical window. The results are shown in Figs.~\ref{fig:Rel} (c) and (d) and mirror those from Figs.~\ref{fig:Rel} (a) and (b). 

In Fig.~\ref{fig:Rel}, our choice of the window of energy in the middle of the spectrum implies infinite temperature. Studies of the dependence of the size of the fluctuations on temperature can also be done~\cite{Santos2010PREb}. The fluctuations are expected to decrease as the temperature increases.

The small fluctuations of the EEV, which happens for chaotic eigenstates, are strong indications that Eq.~(\ref{Eq:thermal}) should hold. But for this to be indeed the case, the initial state needs to probe those chaotic states. We can then single out conditions that guarantee the onset of thermalization: the initial state is highly delocalized, so that equilibration can take place; the initial state has significant overlaps with chaotic eigenstates, that is $E_{\text{ini}}$ falls within the chaotic region of the spectrum; and the width of the energy distribution of the initial state is smaller than or equal to the microcanonical window $\delta E$ \cite{Borgonovi2016}.

\begin{figure}[htb]
\centering
\includegraphics*[width=0.8\textwidth]{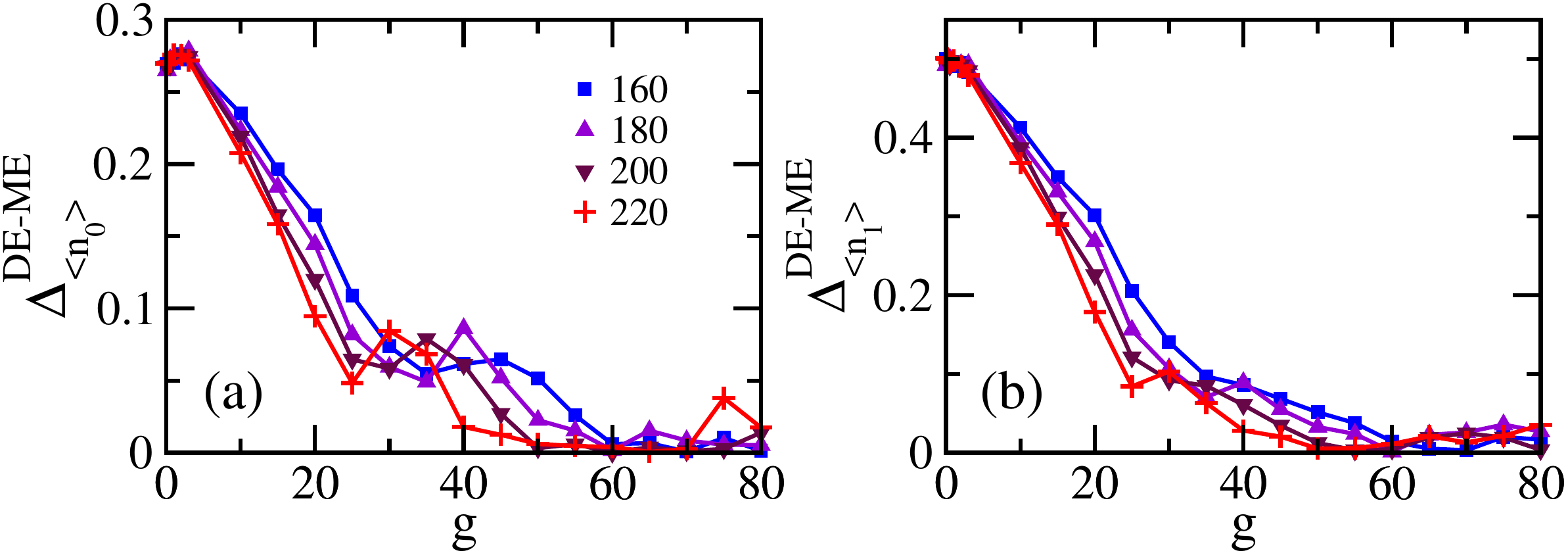}
\caption{Relative difference between the infinite-time average and the microcanonical average as a function of the interaction strength and for different numbers $N$ of atoms (indicated). The initial state is chosen according to Eq.~(\ref{Eq:InitialState}). The microcanonical window is centered at the energy of the initial state, $[E_{\text{ini}} - \delta E, E_{\text{ini}} + \delta E]$ with $\delta E=0.5$.}
\label{fig:DE-ME}
\end{figure}

In Fig.~\ref{fig:DE-ME}, we finally compare the infinite-time average for the initial states chosen according to Eq.~(\ref{Eq:InitialState}) with the microcanonical average. For this, we compute the relative difference,
\begin{equation}
\Delta^{\text{DE-ME}}_{O} = \left| \frac{  O_{\text{DE}} - O_{\text{ME}}  }{  O_{\text{DE}} } \right|.
\end{equation}
The two averages get indeed closer as $g$ and $N$ increase, confirming our expectations that thermalization should take place.

In addition to strong interactions and large numbers of atoms, the energy $E_{\text{ini}}$ of the initial state also plays a role in pushing the system toward thermal equilibrium. The vertical lines in Fig.~\ref{fig:PR3} and Fig.~\ref{fig:EEV} mark the position of $E_{\text{ini}}$. One sees that it moves closer to the middle of the spectrum as $g$ increases. This further contributes to the viability of thermalization. Theoretically, we could also study the dependence of $\Delta^{\text{DE-ME}}_{O}$ on the energy of the initial state for fixed $g$'s and $N$'s~\cite{Torres2013}. Experimentally, we are restricted to the initial states that can be actually prepared.

We chose not to show the results for $\Delta^{\text{DE-ME}}_{\langle n_2 \rangle}$ in Fig.~\ref{fig:DE-ME}. For the selected  initial state only modes 0 and 1 are initially populated, so when $g$ is small, the discrepancy between $\langle n_2 \rangle_{\text{DE}}$ and $\langle n_2 \rangle_{\text{ME}}$ is very large. However, the difference decreases rapidly as the interaction increases and shows results similar to those for $\langle n_0 \rangle$ and $\langle n_1 \rangle$ when $g>20$.

We present in App.~\ref{sec:two-mode} the study of the quantum dynamics for the two-mode model. While this model is insufficient to describe the system, it is interesting to emphasize differences and similarities with the three-model model. We mention two points. (i) Similarly to the three-mode model, with two modes one also finds damping of the oscillations. (ii) Interestingly, with two-modes there is absence of a transition to the quantum chaos regime. In contrast, two-mode model exhibits an excited state quantum phase transition (ESQPT), as expected from it similarity with the Lipkin-Meshkov-Glick Hamiltonian.

\subsection{Quantum chaos and damping: extrapolation to large $N$}

\begin{figure}[htb]
\centering
	\includegraphics*[width=0.5\textwidth]{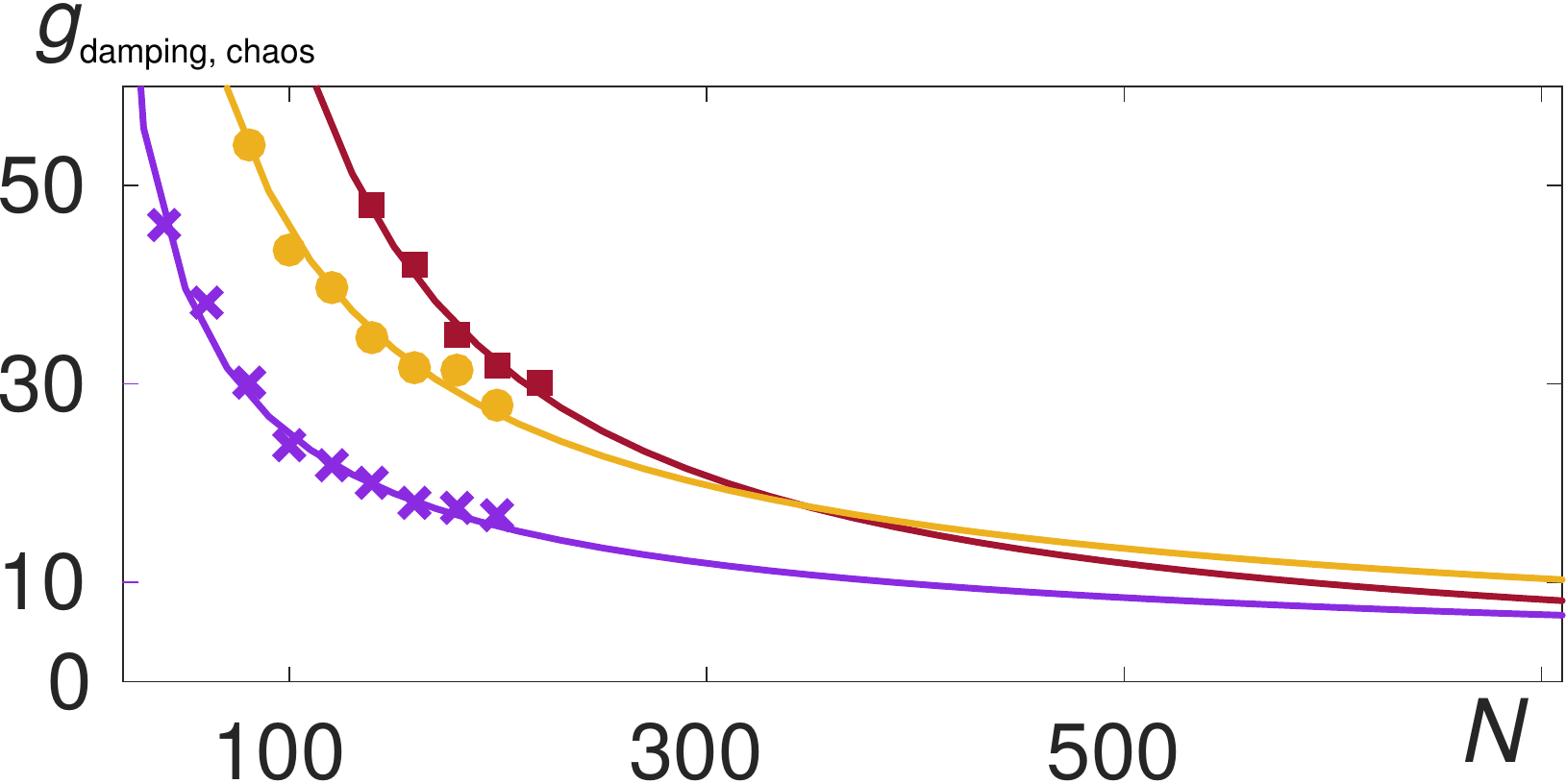}
 	\caption{For each $N$, values of $g$  at which the damping time is smaller than $\tau=2$ (which we name as $g_{\rm{damping}}$) are shown with crosses for the criterion used in Fig.~\ref{fig5} (a) and with circles for the criterion in Fig.~\ref{fig5} (b)]. We also show as a function of $N$, values of $g$ at which the chaos indicator $\beta>0.3$, which we name as $g_{\rm{chaos}}$ and represent with squares.  The solid lines correspond to fittings to the numerical results.  The three curves have the same qualitative behavior. The extrapolation to $N=700$ gives $g\in[8,14]$. }
	\label{fig11}
\end{figure}

We now have the tools to compare the emergence of the damping of the oscillations with the onset of quantum chaos. For this, we choose thresholds for the damping time and chaos indicator $\beta$. For each $N$, we find the values of $g$ at which the damping is so strong that the damping time $\tau$ in Fig.~\ref{fig5} is smaller than 2. We use this convention to get a value for $g_{\rm{damping}}$, which we defined in Sec.~\ref{sec:damping}. Using this convention, we get a set of values of $g_{\rm{damping}}$ as a function of $N$ for the criterion used in Fig.~\ref{fig5} (a) and another one for the criterion used in Fig.~\ref{fig5} (b). The values of $g_{\rm{damping}}$ {\em vs} $N$ are plotted in Fig.~\ref{fig11}. For each $N$, we also obtain the value of $g$ for which $\beta$ in Fig.~\ref{fig:WD} is larger than $0.3$. We call this  $g_{\rm{chaos}}$, as it indicates that the system has already moved away from the integrable point and is approaching the chaotic regime. The behavior of the curves for $g$ {\em vs} $N$ extracted from $\tau$ and from $\beta$ is very similar: the larger the number of atoms is, the smaller the interaction needs to be for damping and chaos.

We note, however, that damping does not require the onset of chaos, as characterized by a Wigner-Dyson distribution. Damping can take place provided we do not encounter an excessive number of degeneracies or commensurate phases. Quantum chaos is a stronger condition to guarantee that not only the system relaxes, but it also reaches an equilibrium described by the Gibbs ensemble. This is why we chose as threshold for the chaos indicator $\beta>0.3$ instead of a value closer to 1.

Similarly to what we did in Fig.~\ref{fig5}, by fitting a curve to each set of data in Fig.~\ref{fig11}, we extrapolate our results to $N=700$, which is the typical number of atoms in the experiments. This leads to a value of $g \sim 10$. Both analysis performed here, based on the damping time and on the approach to chaos,  show that the  strong damping described by the quantum model takes place at larger $g$ than the damping observed experimentally~\cite{Bonneau2018}. 


\section{Conclusion}
\label{sec:ccl}

We have shown that the  three-mode quantum many-body model is a minimal model to qualitatively   describe both the atomic density distribution oscillations and their damping. This behavior is qualitatively similar to the one observed experimentally with a quasi-1D BEC prepared in a coherent superposition of its two lowest motional states~\cite{Bucker2013,vanFrank2014,vanFrank2016}.
 This system is isolated, so it does not include a mechanism for damping through an environment. Yet, one can make a system-environment analogy by viewing the second excited mode, which is essential for the decay of the oscillations, as a minimal environment, and the ground and first excited modes as constituting the system. 

To characterize the observed decay of the oscillations, we employed the exact diagonalization of the many-body Hamiltonian for a number $N$ of atoms ranging from 40 to 220 and a range of the interaction strength $g$. We  showed that the damping time decreases as $g\,N$ increases. The model also undergoes a transition to the quantum chaos regime when $g$ becomes sufficiently strong. This value decreases as $N$ increases. A key finding of this paper is the link established between the decay of the oscillations, the loss of coherence (fragmentation), and the approach to chaos.

The extrapolation of our results to the smallest number of atoms considered in the experiments ($N=700$) reveals that, despite qualitatively reproducing the decay of the oscillations, the many-body three-mode model predicts damping times that are larger than those observed experimentally. We conjecture that this may be due to the  fact that the experimental system is not a true quasi-1D system, but a cigar shaped condensate. For large interactions, phenomena occurring in the elongated direction may be the cause of an extra damping mechanism, which makes the damping time shorter. Whether this mechanism is the twin-atom generation processs described in~\cite{Bucker2011} is out of the scope of this paper and a question to be investigated as an outlook.  

The three-mode model offers a good example for studies of relaxation and thermalization in isolated quantum many-body systems. We have numerically shown that thermalization can indeed take place as $g$ increases. The viability of thermalization is tightly connected with the onset of chaos. We expect that similar results can be found in other  three-mode many-body models, as e.g. three bosonic species with coherent couplings in a trap  or ultracold atoms in three-wells as in Ref.~\cite{WilsmannARXIV}.  The role of the interaction energies that lead to the transfers between modes in our system would be played by the coherent coupling between species in the first model and by the tunneling energies between wells in the second one.  The initial condition in these cases would be a coherent superposition of two of the species for the first model and two of the wells for the second one.  

As a final remark, we mention a new study~\cite{Gallego2017} about the conditions required to prepare an initial state (in general a Hamiltonian protocol) that does not equilibrate, thus introducing the concept of resilience against equilibration. This suggests a link between the area of nonequilibrium quantum dynamics and that of quantum resource theory. In the system studied here, an interesting outlook would be to study the resilience of possible initial coherent states.

\section*{Acknowledgements}
We thank encouraging and fruitful discussions with B. Julia-Diaz, T. Berrada, C. Gogolin, P. Grzybowski and J.F. Schaft. 
We acknowledge the Spanish Ministry MINECO (National Plan 15 Grant: FISICATEAMO No. FIS2016-79508-P, SEVERO OCHOA No. SEV-2015-0522, FPI), European Social Fund, Fundaci\'o Cellex, Generalitat de Catalunya (AGAUR Grant No. 2017 SGR 1341 and CERCA/Program), ERC AdG OSYRIS, ERC advanced Grant QuantumRelax, EU FETPRO QUIC, and the National Science Centre, Poland-Symfonia Grant No. 2016/20/W/ST4/00314.
 M.B.  was  supported  by  the  EU  through  the  Marie Sklodowska Curie grant ETAB (ga no.656530) and by the FWF through the Lise Meitner grant CoPaNeq (M2088-M27).  L.F.S. was supported by the NSF Grant No. DMR-1603418.

\appendix
\section{The two-mode model dynamics and spectrum }
\label{sec:two-mode}

\begin{figure}[htb]
\centering
	\includegraphics*[width=0.65\textwidth]{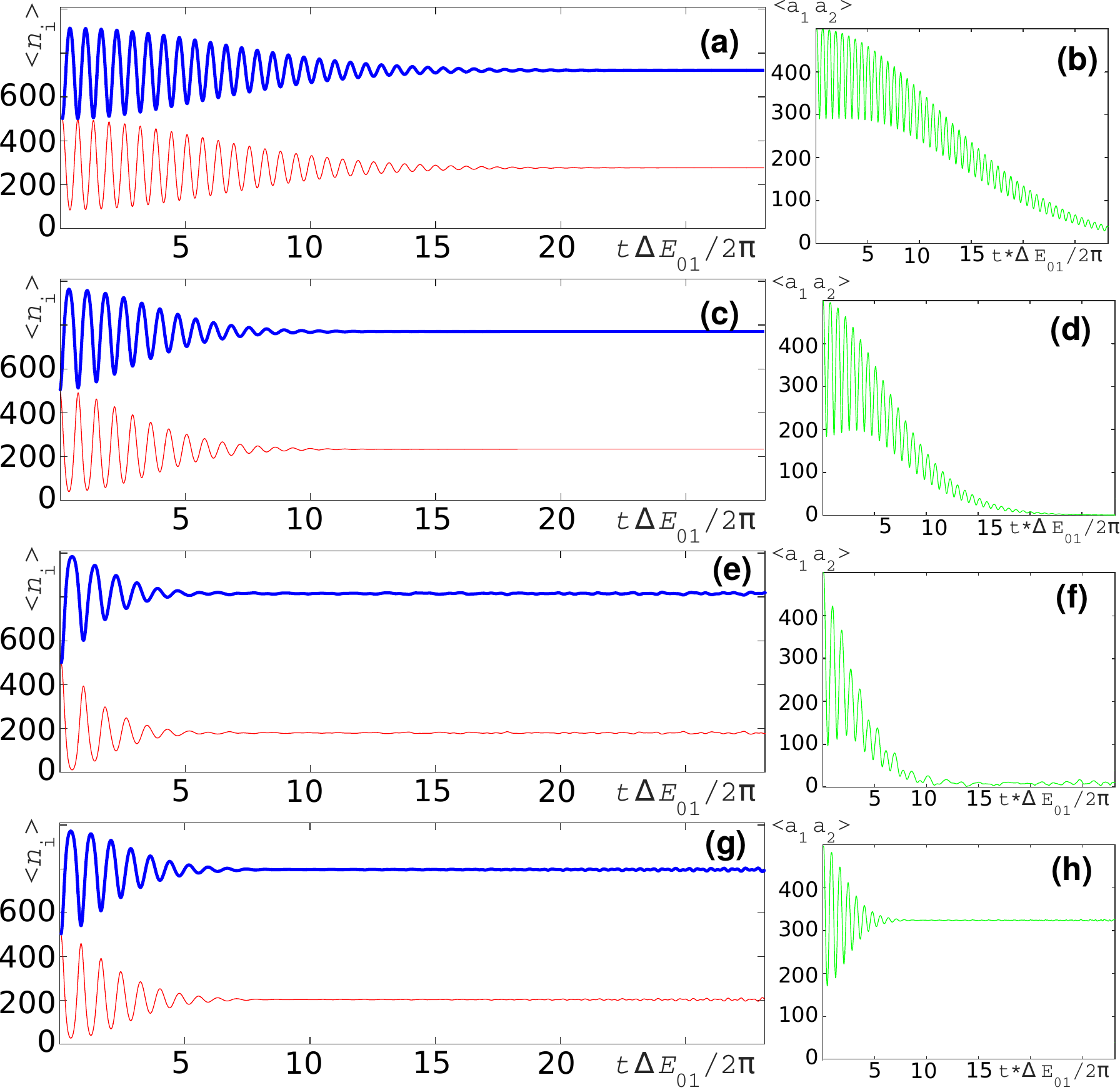}
	\caption{Evolution of the mode average occupations (left column) and the off-diagonal elements of the one body density matrix (OBDM) $\langle a^\dagger_{1} a_{2}\rangle$ (right column)  for the two-mode many-body model for $N=1000$ atoms (same initial state as in Fig.~\ref{fig2}).  On left column, we represent $\langle n_{0,(1)}\rangle $ with a red thin (blue thick) line.  From the top to the bottom panel, the interaction is increased as $g=5,6,6.5$ and 7.     } 
	\label{fig12}
\end{figure}

\begin{figure}[htb]
\centering
	\includegraphics*[width=1.01\textwidth]{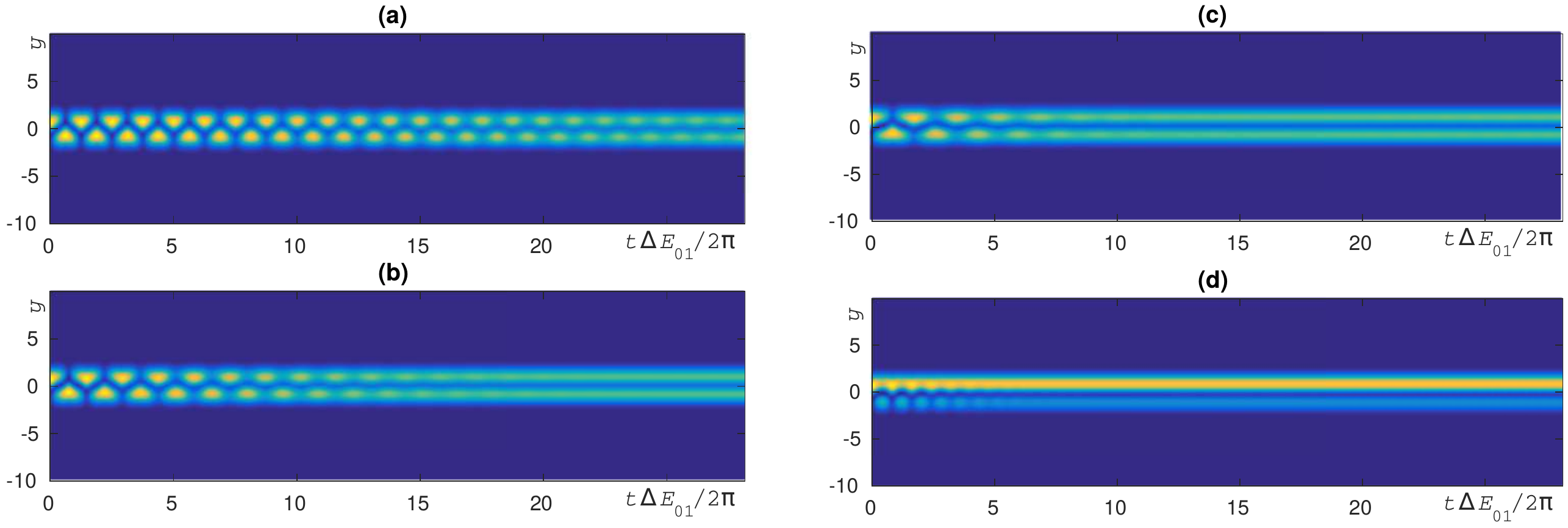}
 	\caption{Evolution of the density profiles for $N=1000$ atoms for the two-mode  many-body model  (same initial state as in Fig.~\ref{fig2}). (a) to (d) correspond to $g=5, 6, 6.5$ and $g=7$, respectively. Damping occurs for similar $g N$ as in the three mode model, but qualitatively the final state is different both to that reached at large times in the three-mode model and in the experiment. }
	\label{fig13}
\end{figure}

If only two modes are considered, we approximate the field operator $\hat \Psi$ describing the condensate by
\begin{equation}
\hat \Psi \simeq  a_0 \psi_0 +  a_1 \psi_1, \label{eq:Ansatz_psi}
\end{equation}
where the $\psi_i$ are the two lower-lying eigenstates of the non-interacting part of the Hamiltonian (taken to be real and normalized to $\int |\psi_i|^2 {\rm d}y = 1$) and the $\hat{a}_i$ are annihilation operators associated with the modes, fulfilling the commutation relation $[ \hat{a}_i, \hat{a}_j^{\dagger}] = \delta_{ij}$. 
Following the approach of~\cite{GarciaMarch2012}, we obtain the effective two-mode Hamiltonian
\begin{eqnarray}
\hat H_\text{2m} &=\frac{ \Delta E}{2} \, ( a_1^\dagger  a_1 -  a_0^\dagger  a_0 )+ \frac{U}{4} ( a_1^\dagger  a_1 -  a_0^\dagger  a_0)^2 +  U_{01} ( a_0  a_1^\dagger +  a_0^\dagger  a_1)^2 , 
\label{eq:hamiltonian} 
\end{eqnarray}
with
\begin{eqnarray}
\Delta E &= E_{01} - (N-1)(U_{00} - U_{11}) , \label{eq:merge} \\
U &= U_{00} + U_{11} - 2 U_{01} , \label{eq:interaction} \quad \text{and} \quad U_{ij} &= \frac{g}{2}  \int d y |\psi_i|^2 |\psi_j|^2  .
\end{eqnarray}
To connect with conventional approaches, let us introduce the operators $ J_x = ( a_0  a_1^\dagger +  a_0^\dagger  a_1)/2$, $ J_y = ( a_0 a_1^\dagger -  a_0^\dagger  a_1)/2i$ and $ J_z = ( a_1^\dagger  a_1 -  a_0^\dagger  a_0)/2$, which satisfy angular momentum commutation relations. We can write  Eq.~(\ref{eq:hamiltonian}) in the spin representation 
\begin{equation}
\hat H_\text{2m} = \Delta E \,  J_z + U  J_z^2 + 4 U_{01}  J_x^2 .
\label{eq:hamiltonianSPIN} 
\end{equation}
In this way we show that this Hamiltonian resembles the bosonic Josephson Hamiltonian, with an additional energy offset between the two modes. The many-body dynamics and damping of the oscillations described with the two-mode model is very different from that described with three-mode model  (see Figs.~\ref{fig2}, \ref{fig3}, and \ref{fig4}).  In Fig.~\ref{fig12} we show the evolution of the mode amplitudes for the two modes and the off-diagonal correlations $\langle a^\dagger_{1} a_{2}\rangle$ for $N=1000$ atoms. The initial condition is an equally weighted coherent superposition of the atoms in the ground and first excited states, $(N_0, N_1) = (500, 500)$ and all relative phases equal to zero.  The first observation is that the fast oscillation observed in the three-mode mode is the only one present in the two-mode model. But more importantly, as $g$ is increased, the two-mode model also shows  damping of the oscillations. This damping is qualitatively different from that observed in the three mode case and in the experiment, as observed from  Fig.~\ref{fig13},  where the corresponding densities are depicted.  First, the final state is different. It is also a fragmented state, but only over the two modes considered. For $N=1000$ we observe that the very quick damping (damping time smaller than two oscillations) occurs also around $g=6.5$, which is of the same order of the one observed for the three-mode model for $N=700$. We note that, for $g>6.5$, the off-diagonal elements   $\langle a^\dagger_{1} a_{2}\rangle$ do not tend to zero anymore.


To better understand the two-mode model, we discuss its Hamiltonian, eigenvalues, and eigenstates. When $U=0$, Eq.~(\ref{eq:hamiltonian}) represents the Lipkin-Meshkov-Glick (LMG) model~\cite{Lipkin1965a}, with the case of $U\neq 0$ being a generalization. This model is integrable and therefore presents no level repulsion. It is also known to exhibit an excited state quantum phase transition (ESQPT). 

In systems with a quantum phase transition, the gap between the ground state and the first excited state closes in the thermodynamic limit. In systems with an ESQPT~\cite{Cejnar2006,Caprio2008}, this crossing occurs together with the clustering of the levels near the ground state and this divergence (peak) of the density of states moves to higher energies as the control parameter increases above the ground-state critical point. Concomitantly, the eigenstates that are very close to the energy of the ESQPT are highly localized leading to the slow evolution of initial states with similar energy~\cite{SantosBernal2015}. 

\begin{figure}[htb]
\centering
\includegraphics*[width=0.8\textwidth]{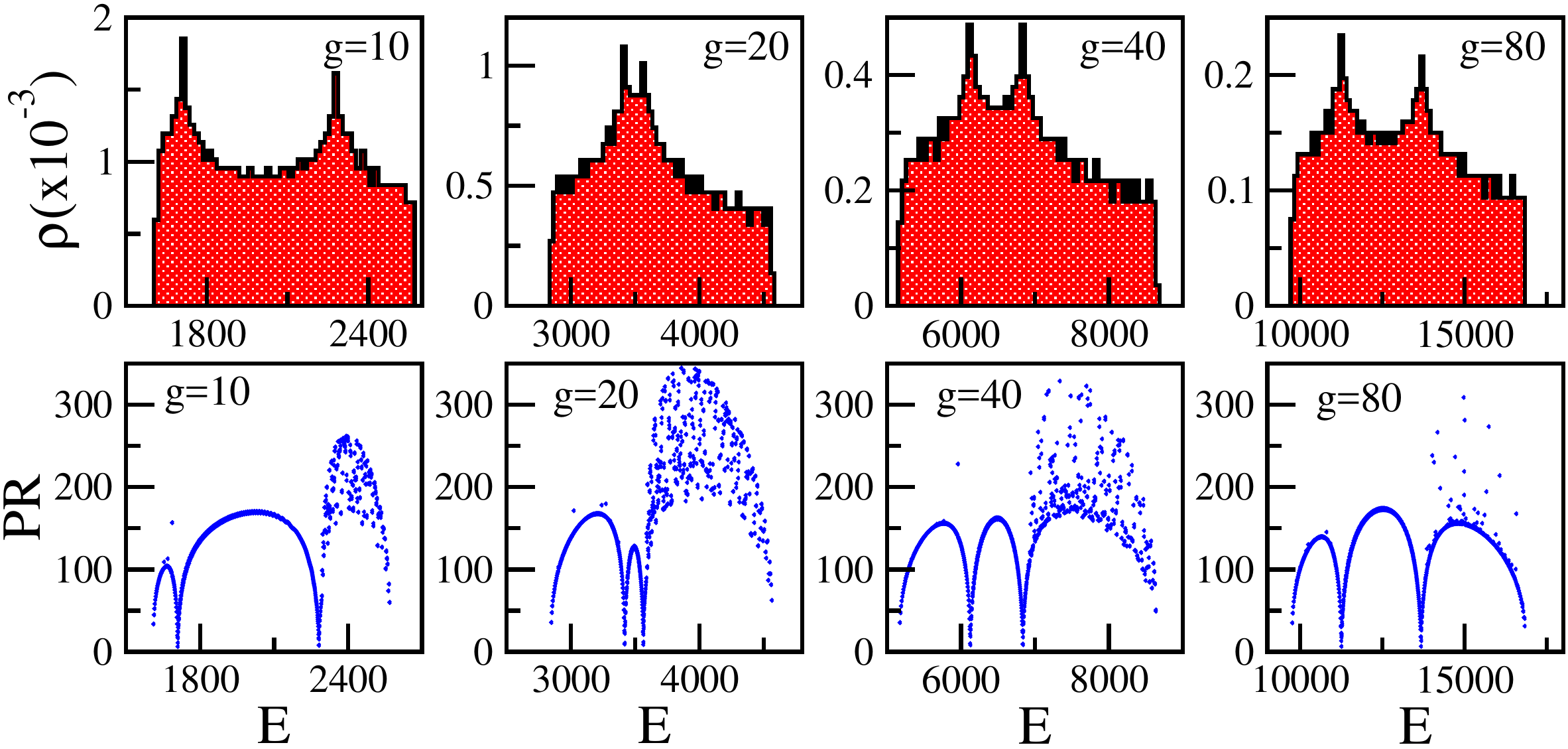}
\caption{(Color online) Density of states $\rho$ (top) and participation ratio (bottom) for the two-mode model for different values of the interaction strength (indicated); $N=1000$. Both parity sectors are included.}
\label{fig14}
\end{figure}

The features of ESQPT for the Hamiltonian (\ref{eq:hamiltonian}) are evident in Fig.~\ref{fig14}. The top panels show results for the density of states, where two peaks are seen. They must be related with two different phase transitions caused by the three competing terms in Eq.~(\ref{eq:hamiltonianSPIN}). They emerge for $g > 3$ and are initially at the borders of the spectrum. We verified numerically that $g>3$ is also the minimum value for which the  two-mode model with $N=1000$ shows damping within the longest simulations we performed (that is a time shorter than $\sim$30 oscillations in terms of $\Delta E_{01}/2\pi$).   As $g$ increases, the two peaks approach each other (compare $g = 10$ and $g = 20$), merge together, and then separate again (compare $g = 40$ and $g = 80$). The peaks merge when only two main competing terms remain in Eq.~(\ref{eq:hamiltonianSPIN}).

The bottom panels of Fig.~\ref{fig14} depict the results for the PR for all eigenstates as a function of energy. Dips in the PR occur at the same energies of the divergences of the density of states (cf. top and bottom panels of the figure). The dips indicate that the eigenstates around the energies of the ESQPT are very localized. 

In summary, the two-mode model is significantly different from the three-mode model. Besides not being chaotic, it exhibits an ESQPT,  which should affect the relaxation process.

\section{Condition for relaxation}
\label{sec:boundsRT}

Here, we discuss briefly the main ingredients of the body of theory which studies relaxation in isolated quantum systems (see e.g. ~\cite{Gogolin2016,Srednicki1996,Srednicki1999,Reimann2008,Linden2009,Linden2010,Short2011,Short2012,Reimann2012,Venuti2013,Zangara2013,Kiendl2017}) to highlight  the connections with our discussion on quantum chaos. To this end, let us denote the evolving state through its density matrix $\rho(t)$, with unitary dynamics dictated by a generic Hermitian Hamiltonian $H$, which is determined by its collection of eigenstates  $\left\{ |\psi_{\nu} \rangle \right\} $ and corresponding eigenenergies $\left\{ E_{\nu}  \right\}$. The Hilbert space is of finite dimension. Let us introduce also the dephased state as 
\begin{equation}
 \omega(\rho_{\text{ini}})=\sum_{\nu} p^{(\nu)}_{\text{ini}} | \psi_{\nu} \rangle\!\langle \psi_{\nu} | ,
 \label{eq:diag}
\end{equation}
where $p^{(\nu)}_{\text{ini}}=\langle \psi_{\nu} |\rho_{\text{ini}} |\psi_{\nu} \rangle$ and $\rho_{\text{ini}}$ is the initial state. When the latter is a pure state, $p^{(\nu)}_{\text{ini}} = |C_{\text{ini}}^{(\nu)}|^2$, as used in Eqs.~(\ref{Otime}) and (\ref{OtimeAve}).  According to Eq.~(\ref{eq:PR}), the PR$_{\text{ini}}$ for the chosen initial state projected in the energy eigenbasis is given by
\begin{equation}
\text{PR}_{\text{ini}} =\frac{1}{\sum_{\nu} | p^{(\nu)}_{\text{ini}}|^2} =\rm{Tr}\left(\omega(\rho_{\text{ini}} )^2\right) .
\end{equation}

If the system relaxes to equilibrium, its long-time average agrees with Eq.~(\ref{eq:diag}), that is
\begin{equation}
\lim_{T\to\infty}\frac{1}{T} \int _0^T dt\rho(t)=\omega(\rho_{\text{ini}}) . 
\end{equation}
Equivalently, the expectation value of an arbitrary observable $O$ tends to 
\begin{equation}
\overline{O} = \rm{Tr}\left(\omega(\rho_{\text{ini}} ) O \right),
\end{equation}
which is the expectation value in the dephased state [see also Eq.~(\ref{OtimeAve})]. As explained in the main text below Eq.~(\ref{OtimeAve}), equilibration requires that the temporal fluctuations around $\overline{O}$ be small and decrease with system size. Under the condition of lack of degeneracies, more precisely absence (or a negligible number) of degenerate level spacings~\cite{Reimann2008,Short2011,Short2012}, it has been shown that the variance of the temporal fluctuations is bounded by
\begin{equation}
 \rm{Var}\left(O\rho\right):=\overline{\rm{Tr}\left[O\left(\rho-\omega(\rho_{\text{ini}})\right)\right]}\le \frac{\| O\|^2}{\text{PR}_{\text{ini}}}.
\end{equation}
This means that for a given observable and under the condition mentioned above, relaxation occurs for highly delocalized initial states. In the context of quench dynamics, highly delocalized initial states emerge in systems  perturbed far from equilibrium and where most eigenstates are strongly delocalized. These conditions are fulfilled by both chaotic and also interacting integrable models, as shown numerically in~\cite{Zangara2013}. This justifies the  sentence from the main text: ``equilibration does not require chaos in the sense of level repulsion, but it needs
highly delocalized eigenstates, delocalized initial states, and not too many degeneracies.''

In Ref.~\cite{Short2011} and others that followed, $\text{PR}_{\text{ini}} $ has been named effective dimension, $d_{\rm eff}(\rho_{\text{ini}} )$, as one understands that is the actual dimension used by the initial state to relax to equilibrium, in contrast with the real dimension of the Hilbert state. If the effective dimension is proportional to the dimension of the Hilbert space, ${\cal D}$, as it is often the case in chaotic systems, one expects that the initial state will thermalize after evolution. 

In connection with the discussion presented here, we note that, recently, the phenomena of equilibration and the time scales required to equilibrate have been related to the quantum phenomena of dephasing in~\cite{Oliveira2018}. In this reference, the authors also estimate the equilibration time scale as  roughly the inverse of the dispersion of the relevant energy gaps. As an outlook we find that such ideas can be investigated with the three-mode model.


\section*{References}

\begin{thebibliography}{10}
\bibitem{footnote}
We note that the shape of the density of states in the three-mode model is very
  similar to that found for the three-orbital Lipkin-Meshkov-Glick
  model~\cite{Wang1998}.
\end{thebibliography}
\providecommand{\newblock}{}

\end{document}